\documentstyle[12pt,epsfig]{article}
 1
\def\as{\alpha_{\rm S}}

\def\citenum#1{{\def\@cite##1##2{##1}\cite{#1}}}
\def\citea#1{\@cite{#1}{}}

\def\as{\alpha_{\rm S}}

\def\D{\Delta}

\def\g{\gamma}

\def\hm{ {-{1\over 2}}  }

\def\o{\omega}

\def\s{\sigma}

\def\({\left(}
\def\){\right)}

\def\citenum#1{{\def\@cite##1##2{##1}\cite{#1}}}
\def\citea#1{\@cite{#1}{}}

\def\l1vt{\vec{l_{1\perp}}}

\def\rt{r_{\perp}}
\def\bt{b_{\perp}}
\def\rt2{$r^2_{\perp}$}
\def\bt2{$b^2_t$}

\def\jol1{$J_0(\,l_{1\perp}\,r_{\perp}\,)$}

\def\citea#1{\@cite{#1}{}}

\def\beq{\begin{equation}}
\def\eeq{\end{equation}}
\def\bea{\begin{eqnarray}}
\def\eea{\end{eqnarray}}

\def\eq#1{{eq.~(\ref{#1})}}

\def\bbbz{{\mathchoice {\hbox{$\sf\textstyle Z\kern-0.4em Z$}}
{\hbox{$\sf\textstyle Z\kern-0.4em Z$}}
{\hbox{$\sf\scriptstyle Z\kern-0.3em Z$}}
{\hbox{$\sf\scriptscriptstyle Z\kern-0.2em Z$}}}}
\def\npb#1#2#3{    {\it Nucl. Phys. }{\bf B#1} (19#2) #3}
\def\plb#1#2#3{    {\it Phys. Lett. }{\bf B#1} (19#2) #3}
\def\prd#1#2#3{    {\it Phys. Rev. }{\bf D#1} (19#2) #3}

\def\prl#1#2#3{    {\it Phys. Rev. Lett. }{\bf #1} (19#2) #3}

\def\zpc#1#2#3{    {\it Z. Phys. }{\bf C#1} (19#2) #3}

\def\sjnp#1#2#3{   {\it Sov. J. Nucl. Phys. }{\bf #1} (19#2) #3}

\def\l{\lambda}
\relax
\begin{document}
\begin{titlepage}
\noindent
 \hfill{ FERMILAB - CONF - 96 - 224 - T}\\[2ex]
\begin{center}

{\Large\bf{GLUON  DENSITY IN NUCLEI}}\\[6ex]
{\large \bf { A. L.
Ayala  F$^{\underline{o}}$ ${}^{a)\,b)}$${}^*$\footnotetext{ ${}^*$ E-mail:
ayala@if.ufrgs.br}
,
 M. B. Gay  Ducati ${}^{a)}$$^{**}$\footnotetext{${}^{**}$
E-mail:gay@if.ufrgs.br}
  and 
 E. M. Levin ${}^{c)\,d)\,\dagger}$
\footnotetext{$^{\dagger}$ E-mail: levin@hep.anl.gov;leving@ccsg.tau.ac.il} 
}} \\[1.5ex]

{\it ${}^{a)}$Instituto de F\'{\i}sica, Univ. Federal do Rio Grande do Sul}\\
{\it Caixa Postal 15051, 91501-970 Porto Alegre, RS, BRAZIL}\\[1.5ex]
{\it ${}^{b)}$Instituto de F\'{\i}sica e Matem\'atica, Univ. 
Federal de Pelotas}\\
{\it Campus Universit\'ario, Caixa Postal 354, 96010-900, Pelotas, RS,
BRAZIL}\\[1.5ex]
{\it ${}^{c)}$ Theory Division, Fermi National Accelerator Laboratory}\\
{\it Batavia, IL 60510 - 0500, USA}
\\[1.5ex]
{\it$ {}^{d)}$ Theory Department, Petersburg Nuclear Physics Institute}\\
{\it 188350, Gatchina, St. Petersburg, RUSSIA}\\[3.5ex]
{\it Talk, given by E.M. Levin at RHIC'96 Summer Study, BNL, LI, July 7 - 19, 1996}\\[
3ex]
\end{center}
{\large \bf Abstract:}
In this talk we present our detail study ( { \it theory and numbers } )
  \cite{AGL} on the shadowing corrections
to the gluon structure functions for nuclei. Starting from rather contraversial
information on the nucleon structure function which is originated by the recent
HERA data, we develop the Glauber approach for the gluon density in a nucleus
based on Mueller formula \cite{MU90} and estimate the value of the shadowing
 corrections in this case. Than we calculate the first corrections to the
 Glauber approach and show that these corrections are big. Based on this 
practical observation we suggest the new evolution equation which takes into
account the shadowing corrections and solve it. We hope to convince you that
the new evolution equation gives a good theoretical tool to treat the
 shadowing corrections for the gluons density in a nucleus and, therefore,
it is able to provide the theoretically reliable initial conditions for the
time evolution of the nucleus - nucleus cascade. The initial conditions 
should be fixed both theoretically and phenomenologically before to
attack such more complicated problems as the mixture of hard and soft 
processes in nucleus-nucleus interactions at high energy or the theoretically
reliable approach to hadron or/and parton cascades for high energy
 nucleus-nucleus interaction.
\end{titlepage}

\section{Introduction.}
The main goal of this talk is to share with you our experience and results
 that we got during the last two years reconsidering the whole issue of the
shadowing corrections  ( SC ) to the gluon density in  nuclei \cite{AGL}.
The title which reflects the key problems that we are going to discuss is:
{ \it `` All ( theory and numbers ) about the SC to gluon density in nuclei"}
 
It is well known that  the gluon density is the most 
important physical observable that governs the physics at high energy
(low Bjorken $x$) in deep inelastic processes \cite{GLR}. Dealing with nucleus 
we have to take into account the shadowing correction (SC) due to 
rescattering of the gluon inside the nucleus, which is the main 
point of interest in this paper. We show that SC can be treated theoretically
in the framework of perturbative QCD (pQCD) and can be calculated using the
information on the behavior of the gluon structure function for the nucleon.

The outline of the talk looks as follows. We start with our motivation
 answering the question why we got interested in the SC for nucleus gluon
 density. In section 3 we will discuss the theory and numerics of the Glauber
 ( Mueller ) approach emphasizing it's theory status and the estimates for the
 SC that came out of it. After short discussion in section 4 the first
 corrections to the Glauber approach we will present what we consider as a right
 way of doing, namely, the new evolution equation that sums all SC
 ( section 5 ). In section 6 we are going to discuss our next steps that we
plan to do in a nearest future, while in section 7 we will give our answer
to the hot question:{ \it and what ?}, trying to collect all problems of RHIC
physics that we will be able to answer using our approach.
\section{Motivation.}
Let us start with a brief summary of the HERA results for the nucleon
 structure functions ( parton densities in a nucleon). The experiment
 \cite{HERA} shows that the deep inelastic structure function $F_2(x,Q^2)$
increases in the region of small $x$ ( at high energies):
$$
F_2(x,Q^2)\,\,\propto\,\,\frac{1}{x^{0.2}}\,\,\,\,\,\,for\,\,\,\,\,\,
10^{-2}\,\,>\,\,x\,\,>\,\,10^{-5}
$$
at large and small ( $Q^2 \,\approx\, 1\,-\,2\,GeV^2$ ) values of the photon
virtualities $Q^2$.

 {\bf At first sight} we can conclude from the analysis
based on the DGLAP evolution equations undertaken through all the world
R\cite{GRV} \cite{MRS} \cite{CTEQ} that :

1. The DGLAP evolution equations work quite well and no other 
ingredients are needed
 to describe all the HERA data.

2. The parton cascade is rather deluted system of partons with small parton -
parton interaction which can be neglected in a first approximation. In other
words we do not need any SC to describe the experimental data.

3. The phenomenological input, namely, the quark and gluon distribution at
initial virtuality $Q^2 = Q^2_0$ can be chosen at sufficiently low
values of $Q^2$ using the backward evolution of the experimental data in
the region of $Q^2 \,\approx\, 4 - 5 GeV^2$. Even more, the craziest
 parameterization that we have seen in our life - the GRV one \cite{GRV}
 does it's job perfectly well, starting with $Q^2_0 = 0.3 GeV^2$ {\bf ?!}.

What we have discussed is moreless common opinion of all experts in DIS and
one can  find  it in many plenary and review talks during the last two years.

{\bf However } we would like to draw your attention to several facts which
do not fit to  this common scheme:

{\bf 1.} The best parameterization of the HERA data is not the solution of the DGLAP
equations but a simple formula \cite{BH}:
$$
F_2(x,Q^2)\,\,=\,\,a\,\,+\,\,m\,\log \frac{Q^2}{Q^2_0}\,\log \frac{x_0}{x}
$$
with $a $\,=\,0.078 ; $m$\,=\,0.364 ; $x_0$\,=\,0.074 ; $Q^2_0$\, =\,0.5
$GeV^2$. It is clear that this simple formula cannot be a solution of the
DGLAP evolution equations.  To make obvious this remark it is enough to
recalculate the gluon structure function from the above expression as 
it has been done in Ref.\cite{BH}. Indeed, $x G(x,Q^2)$ turns out to be equal to
$$ xG(x,Q^2) \,\, =\,\,3\,\,\log \frac{x_0}{x}$$
without any $Q^2$ - dependence within the direct contradiction with
 the DGLAP evolution.

{\bf 2.}
  Using the HERA data we can evaluate the parameter which characterizes the
value of the SC, namely \cite{GLR}
\beq \label{1}
\kappa\,\,=\,\,\frac{ 3 \,\pi\,\as}{ Q^2\,R^2}\,\,xG(x,Q^2)\,\,,
\eeq
where $xG(x,Q^2)$ is the gluon structure function and $R^2$ is the radius 
 of area populated by gluons in a nucleon. The physical meaning of $\kappa$
becomes clear if we rewrite it in the form
$$\kappa\,\,=\,\,x G(x,Q^2)\,\frac{\s(GG)}{\pi\,R^2}\,\,,$$ where $\s(GG)$ 
is the cross section of two gluon interaction in our parton cascade calculated
by Mueller and Qiu \cite{MUQI}, namely,
 $\s(GG)\,\,=\,\,\frac{3 \pi^2 \as}{Q^2}$. The physical meaning of this formula
is the probability of the gluon - gluon interaction inside the parton cascade.
It looks very natural if we compare \eq{1} with the small parameter
 for proton - nucleus interaction. Indeed, the parameter which governs the value
of the Glauber corrections for proton - nucleus interaction $\kappa_{pA}\,=\,A 
\frac{\s(pp)}{\pi R^2_A}$, where $A$ is the number of the constituents
 (nucleons) , $\s(pp)$ is the cross section of the interaction of our
 constituents and $\pi R^2_A$ is the area populated by nucleons. The question
 arises what is the value of $R$ in \eq{1}? 
\begin{figure}[htbp]
\centerline{\psfig{file=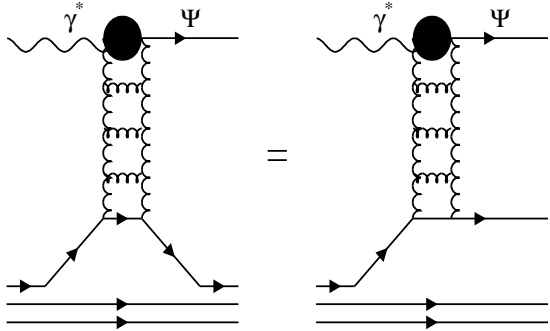,width=90mm}} 
\caption{{\em The $J/\Psi$ production without a) and with b) 
dissociation of the proton.}}
\label{Fig.1}
\end{figure}
  Using the new HERA 
data on photoproduction of J/$\Psi$ meson
\cite{HERAPSI}  we are able to estimate the 
value of $R^2$ in the definition of $\kappa$ (see \eq{1} ).
 To illustrate the point we picture in Fig.\ref{Fig.1} the process of 
J/$\Psi$ photoproduction in the additive quark model (AQM ). We see that 
we have two processes with different slopes ($ B$ )  in $t$
 ( or in $b^2_{\perp}$ 
): the J/$\Psi$ production without ( Fig.\ref{Fig.1}a ) ($B_{el}\,=\,
5\, GeV^{-2}$)  and with ( Fig.\ref{Fig.1}b ) 
( $B_{in}\,=\,1.66 \,GeV^{-2}$ ) dissociation of the  proton.
 The AQM gives us the simplest estimates for the resulting slope ( 
$R^2$ ) in \eq{1} if we neglect any slope from the Pomeron - J/$\Psi$ 
vertex in Fig.\ref{Fig.1}, namely
\beq \label{2}
\frac{1}{R^2}\,\,=\,\,\frac{1}{4}\,\{\,\frac{3}{2 B_{el}} 
\,\,+\,\,\frac{1}{2 B_{in}}\,\}\,\,\approx\,\,\frac{1}{5} \,GeV^{-2}\,\,.
\eeq

\begin{figure}[hbtp]
\centerline{\epsfig{file=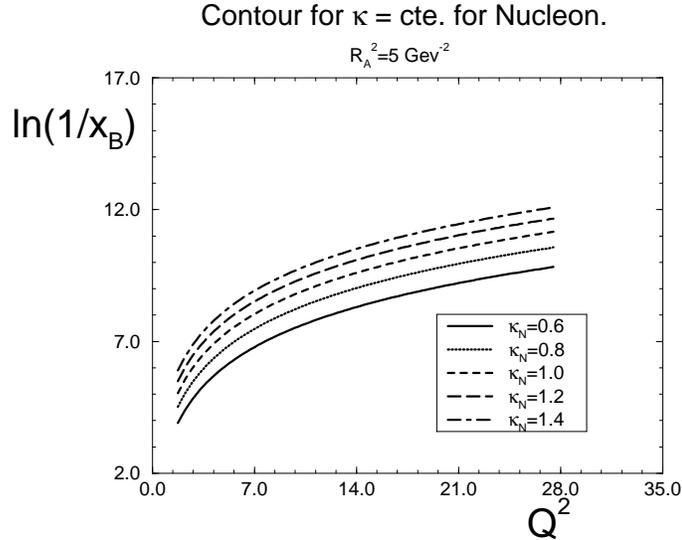,width=90mm}} 
\caption{{\em Contour plot for $\kappa$ for $R^2= \, 5 Gev^{-2}$.}}
\label{Fig.2}
\end{figure}
 Fig.\ref{Fig.2} shows the contour plot for 
$\kappa$ using the GRV parameterization \cite{GRV} for the gluon 
structure function and the value of $R^2 = 5 GeV^{-2}$.
 One can see that $\kappa$ reaches $\kappa$ = 1 at HERA 
kinematic region, meaning shadowing corrections take place.

{\bf 3.}
 The situation looks even more contraversparameterizationial if we plot the average value of
the anomalous dimension 
 $ < \gamma  >\,\,=\,\,\partial \ln( x G(x,Q^2))/\partial \ln Q^2$ in the
GRV parameterization.
\footnote{ We will discuss below the definition of the anomalous dimension
and why this ratio is the average anomalous dimension.} Fig.3 shows two 
remarkable lines: $< \gamma > =1$, where the deep inelastic cross section
reaches the value compatible with the geometrical size of the proton, and
$ < \gamma > \,=\,1/2$, which is the characteristic line in whose vicinity both
the BFKL Pomeron ( see Ref.\cite{BFKL} )  and the GLR equation \cite{GLR} 
should take over the DGLAP evolution equations.We will discuss later what are
 the BFKL and the GLR equations, what we need to know right 
 now, is only the fact
 that both equations give the signal of the new physics. The HERA data passed
 over the second line and even for sufficiently small values of $Q^2$ they 
crossed the first one without any indication of a strange behaviour
near these lines.

\begin{figure}[htbp]
\centerline{\epsfig{file=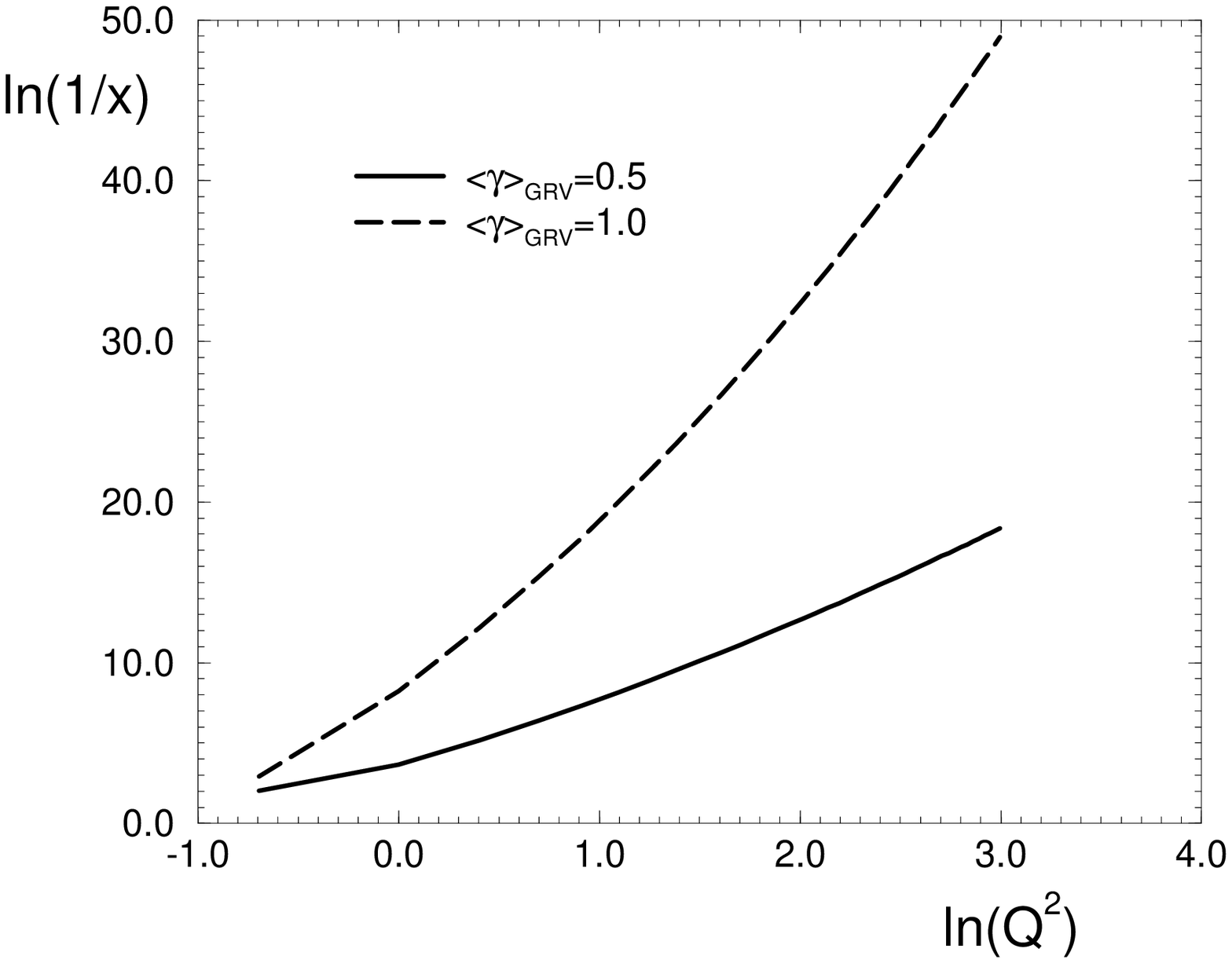,width=90mm}}
\caption{{\em Contours for $ <\gamma > $ = 1 and $<\gamma$ = 1/2.}}
\label{Fig.3}
\end{figure}

Concluding this brief summary of the HERA data and physics behind
them we would like to repeat that  to our taste the situation at HERA looks very
controversial and the statement that the DGLAP evolution works is first but 
not the last  outcome of  the HERA data. On the other hand we have to develop
the new approach to the SC, more general than the GLR one, which will allow us
to give reliable estimates for the SC in the kinematic region to the left of the
line $< \gamma >\,=\,1/2$. This is why we decided to reconsider everything
 that has been known about the SC, trying to forget everything that we knew
 about them, and to start our analisys of the SC from the very beginning.
We also decide to choose the gluon density in a nucleus as a laboratory or
training ground for the new approach to  the problem of the SC.

 We have three reasons for such a choice: (i) the nucleus DIS is easier
 to handle theoretically, as we will show in the main body of our talk;(ii)
the previous analysis of the SC  shows that this is mainly density effect in
 the parton cascade ( see review \cite{LALE} for example ) and we anticipate
larger gluon density for DIS with a nucleus; (iii) the RHIC is coming
 and the gluon density in nuclei will provide the initial condition for
any phenomenological cascades for nucleus - nucleus interaction
 at high energies.

\begin{figure}[htbp]
\centerline{\epsfig{figure=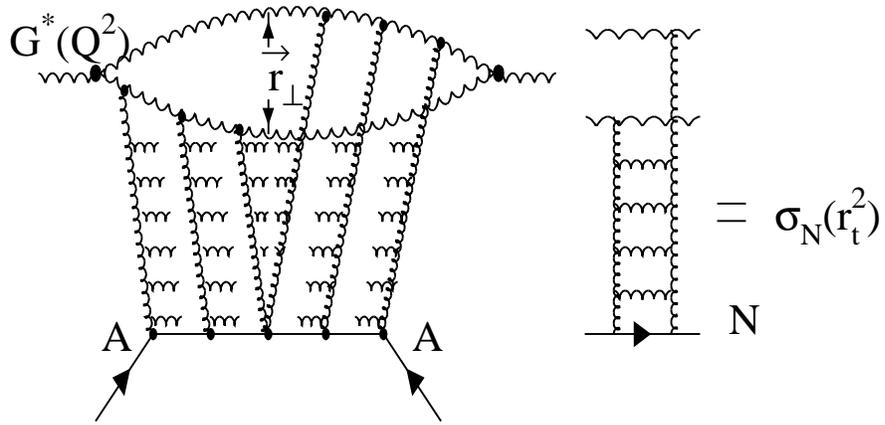,height=100mm}}
\caption{{\em The structure of the parton cascade in the Glauber ( Mueller) 
formula.
 $A$ denotes
the nucleus, N - the nucleon,  $G^{*}(Q^2)$ - the virtual gluon and
$\s_N (r^2_t)$ is the nucleon cross section.}}
\label{Fig.4}
\end{figure}

\section{ The Glauber approach in QCD .}

{\em 3.1 The Mueller formula.}

The idea how to write the Glauber formula in QCD was originally formulated in 
two papers
Ref.\cite{LR87} and Ref. \cite{MU90}. However, the key paper for our problem
is the second paper of A. Mueller who considered the Glauber approach for
the gluon structure function.
Nevertheless, it is easier to explain
the main idea considering the penetration of quark -  antiquark pair,
produced by the virtual photon, through the
target. While the boson projectile is traversing the target, the distance
$r_{\perp}$ between the quark and anti-quark can vary by amount $\D r_{\perp} 
\propto R_{A} \frac{k_{t}}{E}$ where $E$ denotes the energy of the pair in the
target rest frame and $R_{A}$ is the size of the target (see Fig.\ref{Fig.4}).
The quark transverse momentum is $k_{t} \propto 1/r_{\perp}$. 
Therefore
\begin{equation} \label{3}
\Delta \,r_{\perp}\, \propto \,R\,\,\frac{k_t}{E}\,\,\ll\,\,r_{\perp} \, ,
\end{equation}
and is valid if
\begin{equation}
\,r^2_{\perp} \, s \,\,\gg\,\,\,2\,m\,R \, ,
\end{equation}
where $s=2mE$. In terms of Bjorken $x$, the above condition looks as 
follows
\begin{equation}
x\,\,\ll\,\,\frac{1}{2\,m\,R} \, .
\label{4}
\end{equation}
 Therefore the transverse distance between quark and antiquark is a  good 
degree of freedom \cite{LR87}\cite{MU90}\cite{MU94}. As has been shown 
by A.Mueller, 
not only quark - antiquark pairs can be considered in a such way. 
The propagation of a gluon through the target can be treated in a similar
way as the interaction of gluon - gluon pair with definite transverse
separation $r_t$ with the target. It is easy to understand if we remember
 that virtual colorless graviton or Higgs boson is a probe of the gluon
 density. 

The total cross section of the absorption
of gluon($G^*$) with virtuality $Q^2$ and Bjorken $x$ can be written 
in the form:
\beq \label{5}
\sigma^A_{tot}(\,G^*\,)\,\,=\,\,
\int^1_0 d z \,\,\int \,\frac{d^2 r_t}{2 \pi}\,\,
\int\frac{d^2 b_t}{2 \pi}
\Psi^{G^*}_{\perp} (Q^2, r_t,x,z)\,\,
 \sigma_A (x,r^2_t)\,\,\,\,[{\Psi^{G^*}_{\perp}} (Q^2, r_t,x,z)]^* \, ,
\eeq
where $z$ is the fraction of energy 
which is carried by the gluon, $\Psi^{G^*}_{\perp}$ is the wave 
function of the transverse polarized gluon and $\sigma_A (x,r^2_t)$ is the
cross section of the interaction of the $GG$- pair with transverse separation
$r_{t}$ with the nucleus. This cross section can be written in the form:
\beq \label{6}
\s_A(x, r^2_{\perp})\,\,=\,\,2\,\int \,d^2 b_t\, Im a (x,r_{\perp},b_t)\,\,,
\eeq
where $a$ is the elastic amplitude for which we have the $s$-channel unitarity
constraint:
\beq \label{7}
2\,\,Im\,a(x,r_{\perp},b_{t})\,\,=\,\,|a(x,r_{\perp},b_{t})|^2 \,\,+
\,\,G_{in}(x,r_{\perp}, b_{t})\, ,
\eeq
where $G_{in}$  is the contribution of all the 
inelastic processes. Let us recall that two terms in \eq{7} have different 
physical meaning: the left hand side and the first term in the right hand side
describe the interference between the incoming plane wave and outgoing spherical
wave which amplitude is the elastic scattering amplitude ($a$). These two terms
cannot be calculated using a classical approach or simple Monte Carlo - like
 model. The Quantum Mechanics of the interaction is mostly absorbed in
 these two terms while the last term has a simple probabilistic meaning, namely,
the probability of any inelastic interactions, and can be treated almost
 classically and, for certain, in the probabilistic way, for example
 in Monte Carlo-like models. The unitarity establishes the correlation
 between two unknowns $a$ and $G_{in}$ and has the general solution:
\beq \label{8}
a (x,r_{\perp},b_t)\,\,=
\,\,i\,\{\,\,1\,\,-\,\,e^{-\,\frac{1}{2}\,\Omega(x, r_{\perp},b_t)}\,\,
\}\,\,;
\eeq
$$
G_{in}(x,r_{\perp},b_t)\,\,=\,\,1\,\,-\,\,e^{-\, \Omega (x,r_{\perp},b_t)}\,\,.
$$
One can see that $\Omega$ has a simple physical meaning, namely
$e^{- \Omega}$ is the probability that $GG$-pair has no inelastic
 interaction during the passage through the target.
The opacity $\Omega$ is an arbitrary real function, which can be specified
only in more detail theory or model approach than the unitarity constraint.
One of such specific model is Glauber approach or Eikonal model. 

However,
before we will discuss this model let us make one important remark on the
strategy of the approach to the SC. We are trying to built a model or theory
for the total cross section ( or for the gluon structure function ) not
because the SC should be the strongest one in this particular observable, but
because if we will be able to calculate opacity $\Omega$ we will have the theory
or model for all inelastic processes. Indeed, using AGK cutting rules \cite{AGK}
we can calculated any inelastic process, if we know $\Omega$, in accordance
 with the $s$-channel unitarity. It is worthwhile mentioning that the inverse
procedure does not work. If we know the SC in all details for a particular
inelastic process, say for the inclusive production, we cannot reconstruct
all other process and the total cross section in particular.

Now, let us built the Glauber approach. First, let us assume that $\Omega$ is
 small ($\Omega\,\,\ll\,\,1$ )
 and it's $b_t$ dependence can be factorised as $\Omega\,=\,\widetilde{
\Omega}(x,r_{\perp}) \,S(b_t)$ with the normalization: $\int \,d^2 b_t \,S(b_t)
\,=\,1$.
Expanding \eq{8} and substituting it in \eq{6}, one can obtain:
\beq \label{9}
\s_A(x,r_{\perp})\,\,=\,\,\widetilde{\Omega}(x,r_{\perp})
\eeq
At small $\Omega$ the cross section of the deep inelastic process with a nucleus
is proportional to the number of nucleons in a nucleus ($A$), namely,
$$ \s_A (x, r_{\perp})\,\,=\,\, A\,\s_N(x,r_{\perp}).$$
To calculate $\widetilde{\Omega}$ we need to substitute everything in \eq{5}
and use the formula for
  $\s_A (\,G^*)\,\,=\,\,\frac{4 \pi^2}{Q^2} x G_A (x,Q^2)$ as well as the
expression for the wave function of the $GG$- pair in the virtual gluon probe.
Such calculations has been done in Ref.\cite{MU90} and we recapture here
the result ( see for example Ref. \cite{AGL} for more details ):
\beq \label{10}
\widetilde{\Omega}\,\,=\,\,A\,\s_N (x,r_{\perp} )\,\,=\,\,\frac{ 3\, \pi^2}{4}
\,\,r^2_{\perp}\,xG(x,\frac{4}{r^2_{\perp}})\,\,.
\eeq
The Glauber (eikonal ) approach is the assumption that
 $\Omega \,=\,\widetilde{\Omega}\,S(b_t)$ with $\widetilde{\Omega}$ of
\eq{10} not only in the kinematic region where $\Omega$ is small but
 everywhere. From the point of view of the structure of the final state
 this assumption means that the rich typical inelastic event was modeled
as a sum of the diffraction dissociation of $GG$ - pair plus uniform in rapidity
distribution of produced gluons. For example, we neglected in the Glauber
    approach all rich structure of the large rapidity gap events including
 the diffractive dissociation in the region of large mass.

Substituting everything in \eq{10} and \eq{5} and using the wave function
calculated by Mueller in Ref.\cite{MU90} we obtain the Glauber (Mueller) formula
for the gluon structure function:
\beq \label{MF}
x G_A(x,Q^2) = \frac{4}{\pi^2} \int_{x}^{1} \frac{d x'}{x'} 
\int_{\frac{4}{Q^2}}^{\infty} \frac{d^2 r_t}{\pi r_{t}^{4}} 
\int_{0}^{\infty} \frac{d^2 b_t}{\pi}  2
\left\{ 1 - e^{\hm \s_{N}^{GG} ( x^{\prime},r^2_t ) S(b^2_t) } \right\}
\eeq
 
It is easy to see that the first term in the expansion of \eq{MF}
with respect to $\s$ gives the DGLAP equation in the region of small $x$.

To calculate the profile function $S(b_t)$ we make the usual assumption that
$<b^2_t>_N$ in the interaction of $GG$ - pair with the nucleon is much smaller
 than the nucleus radius ( $ < b^2_t >\,\,\ll\,\,R_A$. Therefore,  $S(b_t)$ 
can be expressed through the nucleon wave function in a nucleus, namely
\beq \label{11}
S_A(q_z,b_t)\,\,=\,\,\int\,\,d z_1 \,\,e^{i\,q_z \,z_1}\,\Psi_A (z_1,b_t;r_2,...
r_i,r_A) \Psi^*(z_1,b_t;r_2,...r_i,r_A)\,\prod^A_{i=2}\,d^3\,r_i\,\,,
\eeq
where the wave function is normalized as
\beq \label{12}
\int \,\Psi_A (z_1,b_t;r_2,...r_i,r_A) \Psi^*(z_1,b_t;r_2,...r_i,r_A)\,
\prod^A_{i=1}\,d^3\,r_i\,\,=\,\,A\,\,.
\eeq
Assuming that there is no correlation between nucleons in a nucleus and
 the simple Gaussian form of a single nucleon wave wave function we derive
the Gaussian parameterization for $S(b_t)$, namely
\beq \label{13}
S_A(q_z,b^2_t)\,\,=\,\,\frac{A}{\pi\,R^2_A}\,e^{ -\, \frac{b^2_t}{R^2_A}\,-\,
\frac{R^2_A}{4}\,q^2_z}\,\,,
\eeq
where the mean radius $R^2_A$ is equal to
$$ R^2_A\,\,=\,\,\frac{2}{5}\,R^2_{WS}$$
and $R_{WS}$ is the size of the nucleus in the Wood-Saxon parameterization. We
choose $R_{WS} \,=\,r_0 \,A^{\frac{1}{3}}$ with $r_0\,=\,1.3\,fm$ in
all our calculation. We are doing all calculation in the rest frame of
 the nucleus where we can neglect the change of energy for the recoil nucleon
 in the nonrelativistic theory for the nucleus. Indeed, its energy is
 $E_{p'}\,=\,m\,+\,\frac{q^2}{2m}$ and $\frac{q^2}{2 m}\,\ll\,q_z$. At high
energy ( small $x$ ) we can neglect also the $q_z$-dependance (
 see Ref.\cite{AGL} for details).

Using Gaussian parameterization for $S(b_t)$ ( see \eq{13} )  we can
take the integral over $b_t$ and obtain the answer ($N_c = N_f = 3$)
\bea
x G_A(x,Q^2) = \frac{2 R_{A}^{2}}{\pi^2} \int_{x}^{1} \frac{d x'}{x'} 
\int^{\frac{1}{Q^2_0}}_{\frac{1}{Q^2}}  \frac{d r_t^2}{ r_{t}^{4}} 
\left\{ C + ln(\kappa_{G} ( x', r_{t}^{2})) 
E_1 (\kappa_{G} ( x', r_{t}^{2}))  \right\}
\label{14}
\eea
where $C$ is the Euler constant and $E_1$ is the exponential integral
(see Ref.\cite{r20} Eq. {\bf 5.7.11}) and
\beq
\kappa_{G} ( x', r_{t}^{2}) = \frac{3 \as A \pi r^2_t}{2 R_{A}^{2}}
x' G_{N}^{DGLAP} (x', \frac{1}{r^2_t} )
\label{kapa}
\eeq
The \eq{MF} is the master equation of this section and it gives a way to
estimate the value of the SC. We would like to stress 
 that we have only  adjusted the approach of Ref. \cite{MU90} to
the rescattering in a nucleus. It means that we did nothing except that we
share the responsibility with A. Mueller for \eq{MF}.

\begin{figure}[htbp]
\centerline{\epsfig{figure=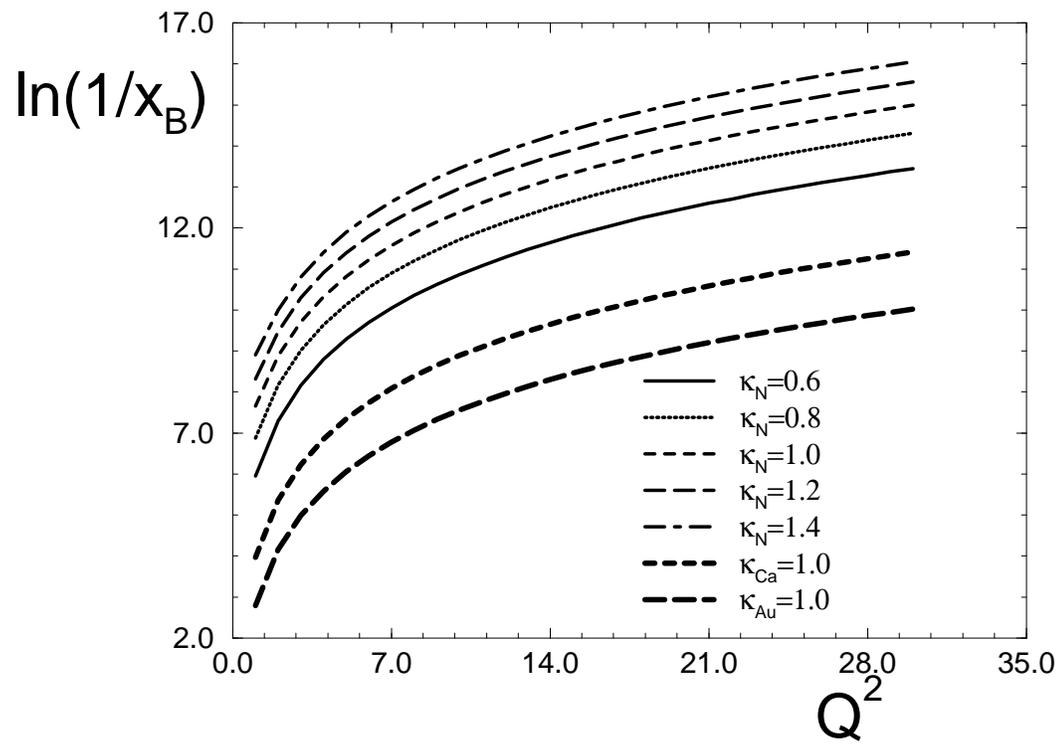,width=140mm}}
\caption{\em The contours of $\kappa$ for Nucleon, Ca and Au. }
\label{kap}
\end{figure}

One can see that the Mueller formula of \eq{MF} depends only on $\kappa$.
If $\kappa$ is small ( $\kappa \,\ll\,1$ ), we can expand \eq{14} and obtain
the DGLAP evolution equation for the gluon structure function. 
If $\kappa\,\gg\,1$, we can use the asymptotic formula for $E_1$ and
obtain:
$$
xG_A(x,Q^2)\,\,=\,\,
 \frac{2 R_{A}^{2}}{\pi^2} \int_{x}^{1} \frac{d x'}{x'} 
\int^{\frac{1}{Q^2_0}}_{\frac{1}{Q^2_0(x')}}  \frac{d r^2_{\perp}}{ r^4_{\perp}} 
\left\{ C + ln(\kappa_{G} ( x', r^2_{\perp} ) \right\}\,\,,
$$
where $Q^2_0(x')$ is the solution of the equation:
\beq \label{15}
\kappa_G (x',r^2_t\,=\,\frac{1}{Q^2_0(x')})\,\,=\,\,1\,\,.
\eeq
  In Fig.\ref{kap} are plotted the contours of $\kappa$ for a nucleon target
   that give 
an idea in which kinematic region we expect big SC.

{\em 3.2  Theory status of the Mueller formula.}

In this section we shall recall the main assumptions that have been made 
to obtain
the Mueller formula.

1. The gluon energy ($x$) should be  high (small) enough to satisfy 
\eq{1} and  \,\,\,\,$\as ln(1/x) \leq 1$. The last condition means that we 
have to assume
the leading $ln (1/x)$ approximation of perturbative QCD for the nucleon gluon
structure function.

2. The DGLAP evolution equations hold in the region of small $x$ or, in other
words, $\as ln(1/r^2_{\perp}) \leq 1$. One of the lessons from HERA data is the
fact that the GLAP evolution can describe the experimental data.

These two assumptions mean that we describe the gluon  emission in so called
 Double Log Approximation ( DLA) of perturbative QCD, or in other words, we
extract from each Feynman diagram of the order $\as^n$ the contribution
of the order $(\,\as\,\ln1/x\,\ln Q^2/Q^2_0\,)^n$, neglecting all other
contributions of the same diagram. In terms of the DGLAP evolution, we
have to assume that the DGLAP evolution equations describe the gluon
 emission in the region of small $x$. However, the first assumption
is very important for the whole picture, since it
 allows us to treat successive rescatterings as independent
 and simplifies all formulae 
 reducing the problem to an eikonal picture of the
classical propagation of a relativistic particle with high energy ($ E \gg 
\mu^{-1}$, where $\mu$ is the scattering radius in the nuclear matter) through 
the nucleus.
The second one simplifies calculations but we can consider
 the BFKL evolution \cite{BFKL} instead of the DGLAP one.  

3. Only the fastest partons ($GG$ pairs) interact with the target.
This assumption is an artifact of the Glauber approach, which looks strange in
 the parton picture of the interaction. Indeed, in the parton model  
we rather  expect that all partons not only the fastest ones should interact
 with the target. 
In the next
 section we will show that corrections to the Glauber approach 
due to the interaction of slower partons are essential in QCD too.

4.  There are
no correlations (interaction) between partons from the different parton cascades
(see Fig.4 ).  This assumption means that even the interaction of 
the fastest $GG$-pair was taken into account in the Mueller formula only
approximately and we have to assume that we are dealing with large number 
of colours to trust the Mueller formula. Indeed, it has been proven that
correlations between partons from different parton cascades lead to 
corrections to the Mueller formula of the order of $1/N^2_c$, where $N_c$ is the
number of colours ( see Ref.\cite{AGL} and references therein for detail
discussions on this subject).

5. There are no correlations between different nucleons in a nucleus.

6. The average $b_t$ for $GG$ pair-nucleon interaction is much smaller than
$R_A$.

The last two are usual assumptions to treat nucleus scattering. We
have used the specific Gaussian parameterization for $b_t$ dependence.
Also, one can easily generalize our 
formula in more general case, as Wood-Saxon parameterization \cite{r33}.

{\em 3.4 The modified Mueller formula.}

The next step of our approach is to give an estimate of the SC using 
the Mueller formula. However, before doing so, we have to study how well works
the DLA of perturbative QCD which was heavily used in the derivation of
 the Mueller formula. Let us recall that the solution of the DGLAP evolution
equations can be easily found in the moments space. For any function $f(x)$
 we define the moment $f(\o)$ as
$$
f(\o)\,\,=\,\,\int^1_0\,\,d x x^{\o}\,f(x)\,\,.
$$
Note that the moment variable $\o$ is chosen such that the $\o=0$ moment
measures the number of partons, and the moment $\o=1$ measures their momentum.
An alternative moment variable $N = \o - 1 $ is often found in the literature.
The $x$-distribution can be reconstructed by considering the inverse
 Mellin transform, which for the gluon distribution reads:
\beq \label{MT}
x G(x, Q^2)\,\,=\,\,\frac{1}{2 \,\pi\,i} \,\int_C\,d\,\o\,\,
 g_{in}(\o,Q^2_0)\,\,
e^{\o\,\ln(1/x)\,\,+\,\,\g(\o)\,\ln(Q^2/Q^2_0)}\,\,,
\eeq
where the contour of integration $C$ is taken to the right of all
 singularities and function $g_{in}$ is defined by the initial gluon
 distribution at $Q^2 = Q^2_0$. The anomalous dimension $\g(\o)$
has to be calculated in perturbative QCD and can be written in the form:
\beq \label{GAMMA}
\g(\o)\,\,=\,\,\frac{\as N_c}{\pi}\cdot\frac{1}{\o}\,\,+\,\,\frac{2 \as^4 N^4_c
\zeta(3)}{\pi^4}\cdot\frac{1}{\o^4}\,\,+\,\,O(\frac{\as^5}{\o^5})\,\,+\,\,O(\as)
\,\,.
\eeq
In  the {\bf DLA} we take only the first term of this series, namely,
$$
\g^{DLA}\,\,=\,\,\frac{\as N_c}{\pi}\cdot\frac{1}{\o}\,\,;
$$
In the {\bf  BFKL} evolution equation all terms of the order
 $( \frac{\as}{\o})^n $ have to be taken into account. They generate the BFKL
 anomalous  dimension of the form:
$$
\g^{BFKL}(\o)\,\,=\,\,\frac{\as N_c}{\pi}\cdot\frac{1}{\o}\,\,+\,\,
\frac{2 \as^4 N^4_c
\zeta(3)}{\pi^4}\cdot\frac{1}{\o^4}\,\,+\,\,\sum^{\infty}_{n=5}
c_n\,(\frac{\as}{\o})^n\,\, |_{\o\,\rightarrow \o_L}\,\rightarrow
\,\,\frac{1}{2}\,\,+\,\,\sqrt{\frac{\o \,-\,\o_L}{\Delta}}\,\,,
$$
where $\g^{BFKL}(\o=\o_L)\,=\,1/2$. The main qualitative property of 
the BFKL anomalous dimension is the fact that it cannot exceed the value 1/2.

The momentum conservation means that $\g(\o= 1)\,=\,0$.
None of the DLA  or the BFKL anomalous dimension satisfies this equation,
because they give the good approximation to the full anomalous dimension only
in the region of small values of $\o$ or, in other words, in the region of
 small $x$.

The DLA anomalous dimension leads to the simple evolution equation:
\beq \label{DLA}
\frac{\partial^2 x G(x,Q^2}{\partial \ln(1/x)\,\partial \ln Q^2}\,\,=\,\,
\frac{\as\,N_c}{\pi}\,xG(x,Q^2)\,\,.
\eeq
 Now let us estimate how well works the DLA. In all our numerical estimates 
we use the GRV parameterization \cite{GRV} for the nucleon gluon distribution,
which describes all available experimental data quite well, including recent
HERA data at low $x$.  Moreover, GRV is suited for our purpose
 because
(i) the initial virtuality for the GLAP evolution is small
 ($Q_0^2 \approx 0.25GeV^2$) and we can discuss the contribution of the 
large distances in MF
having some support from experimental data;
 (ii) in this parameterization the most
essential contribution comes from the region where $\alpha _s ln Q^2 \approx 1$
and $\alpha _s ln 1/ x \approx 1$. This allows the use of the double 
leading log
approximation of pQCD, where the MF is proven \cite{MU90}. It should be also
stressed here, that we look at the GRV parameterization as a solution of
 the  DGLAP
evolution equations, disregarding how much of the SC has been taken into
 account in this parameterization in the form of the initial gluon distribution.

 However, in spite of the fact that the GLAP evolution in the GRV
 parameterization starts from very low virtuality ( $Q^2_0 \,\sim\,0.25 GeV^2$)
it turns out that the DLA still does not work quite well in the accessible
kinematic region ($Q^2\,> 1 GeV^2, x \,>\,10^{-5}$). 
To illustrate this statement we plot in Fig.5 the ratio:
$$
\frac{< \frac{\as N_c}{\pi} >}{\frac{\as N_c}{\pi}}\,\,=\,\,\frac{
\frac{\partial^2 x G^{GRV}(x,Q^2)}{\partial \ln(1/x)\,\partial \ln Q^2}
}{ xG^{GRV}(x,Q^2)}\,\,.
$$
This ratio is equal to 1 if the DLA holds. From Fig.5 one can see that this 
ratio is rather around 1/2 even at large values of $Q^2$.
\begin{figure}[htbp]
\centerline{\epsfig{figure=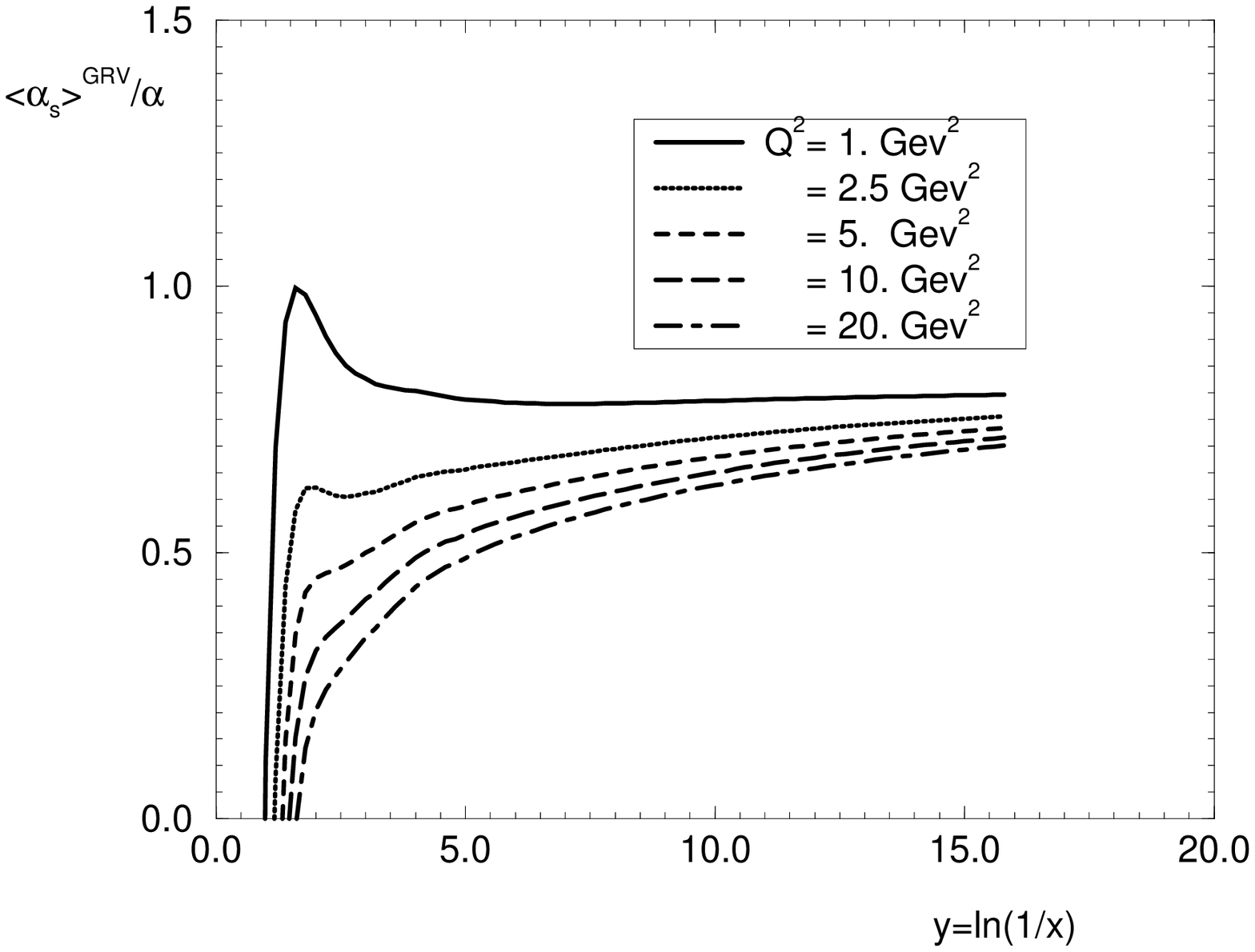,width=90mm}}
\caption{{\em The ration $\frac{< \as >}{\as}$ for different values of $Q^2$
in the GRV parameterization}}
\label{avalpha}
\end{figure}

We can understand why the corrections to the DLA is so big modeling the 
complicated expression for $\g$ of \eq{GAMMA} by simple formula \cite{EKL}
\footnote{ We are very grateful to Yu. Dokshitzer for enlighting discussions
on this problem during  the RHIC'96  Workshop}:
\beq \label{APRGAM}
\g(\o)\,\,=\,\,\frac{\as N_c}{\pi} \cdot \,\{\,\,\frac{1}{\o} \,\,-\,\,1\,\,\}
\,\,.
\eeq

Eq. (\ref{APRGAM}) has correct the DLA limit at small $\o$ and it satisfies 
the momentum conservation ( $\g(\o=1)\,=\,0$). The typical values of $\o$
in all available parametrizations,even in the GRV , which is the closest
 to the DLA, is $< \o > \,\approx\, 0.5 $. Therefore, we have about 50\%
correction to the DLA.  Therefore, the DLA cannot provide a reliable estimates
for the gluon structure function.

On the other hand, our
master equation (see \eq{MF}) is proven in  DLA. Willing to develop a
realistic approach in the region of not ultra small $x$ ($x \,>\,10^{-4})$
we have to change our master equation ( \eq{MF} ). We suggest to 
substitute  the full DGLAP kernel ( the full expression of \eq{GAMMA} )
in the first term of the r.h.s. This procedure gives
\bea
x G_A(x,Q^2)\,\, & = &\,\, x G_A(x, Q^2)(\,\eq{MF}\,) \,\,+\,\,
A x G^{GRV}_N (x,Q^2)\,\,
\nonumber \\
& - &\,\,A\,\frac{\as N_c}{\pi} \,\int^1_x \,\int^{Q^2}_{Q^2_0} \,\,
\frac{d x'}{x'}\,
\frac{ d Q'^2}{Q'^2} \,x' G^{GRV}_N (x',Q'^2)\,\,.
 \label{FINANS}
\eea
 The above equation includes also $A x G^{GRV}_N (x, Q^2_0)$ as the initial
condition for the gluon distribution and gives $A x G^{GRV}_N (x, Q^2)$ as the
 first term of the expansion with respect to $\kappa_G$. Therefore, this
equation is an attempt to include the full expression for the anomalous
 dimension for the scattering off each nucleon, while we use the DLA to 
take into account all SC. Our hope, which we will confirm by numerical
calculation, is that the SC  are small enough for $x \,>\,10^{-3}$ and
we can be not so careful in the accuracy of their calculation in this kinematic
 region. Going to smaller $x$, the DLA becomes better and \eq{FINANS} tends
 to our master equation (\ref{MF}).

{\em The gluon structure function for nucleon (A = 1 ).}

\begin{figure}[htbp]
\centerline{\epsfig{figure=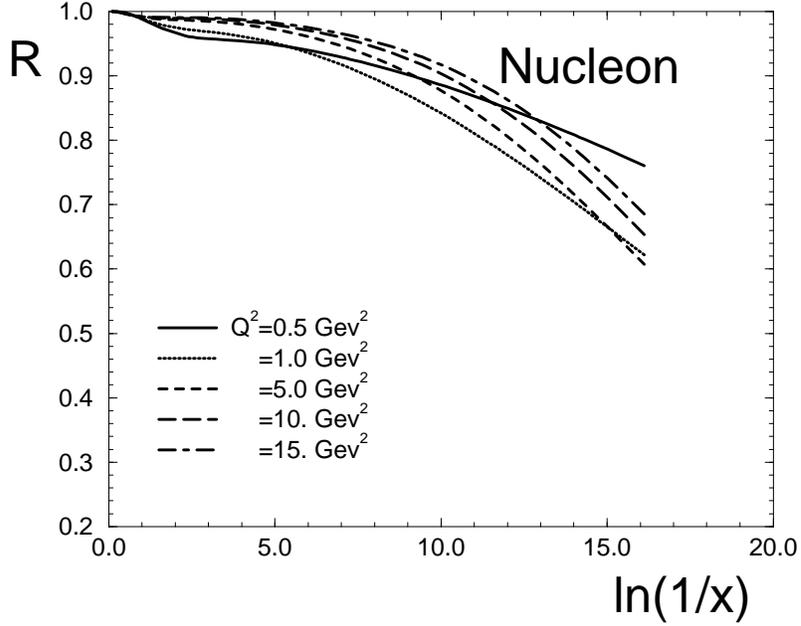,height=90mm}}
\caption{\em The SC  for nucleon (A=1) as a function of $ln(1/x)$ and $Q^2$,
 where ratio $R_1$ compares $xG^A $ with $xG\,\, (GRV)$ distribution. }
\label{r1n}
\end{figure}

In this subsection we are going to check how \eq{FINANS} describes the
gluon structure function for a nucleon, which is our main ingredient in the 
Mueller formula.  We calculate first  the ratio
\beq \label{RN}
R^N_1\,\,=\,\,\frac{xG^A(x,Q^2) (\eq{FINANS})}{x G^{GRV}_N (x,Q^2)}\,\,,
\eeq
for $A=1$, which is shown in Fig.\ref{r1n}. From this ratio we can see the
general behavior of the SC as a function of $ln(1/x)$ and $Q^2$. In the 
region of the HERA data, $3 \, < \, ln(1/x) \, < 10$, and $Q^2 > 2 \, 
GeV^2$\cite{HERA}, the
SC are not bigger than $15 \, \%$. The SC give 
a contribution bigger than $20 \, \%$
 only at very small value of $x$, where we have no experimental data.

In the semiclassical approach (see \cite{GLR}), the nucleon 
structure function is supposed to have  $Q^2$ and $x$ dependence as 
\beq \label{SEMICLAS}
x G^N(x,Q^2)\,\,\propto\,\,\{Q^2\}^{< \gamma >}\,\,
\{\frac{1}{x}\}^{ < \omega >}\,\,.
\eeq

We can calculate both exponents using the definitions
\beq
< \omega >\,\,=\,\,\frac{\partial \ln (x G^N (x,Q^2))}{\partial \ln(1/x)}\,\,.
\label{omega}
\eeq

\beq 
< \gamma >\,\,=
\,\,\frac{\partial \ln (x G^N(x,Q^2))}{\partial \ln (Q^2/Q^2_0)}\,\,;
\label{gama}
\eeq 

\begin{figure}[htbp]
\centerline{\epsfig{figure=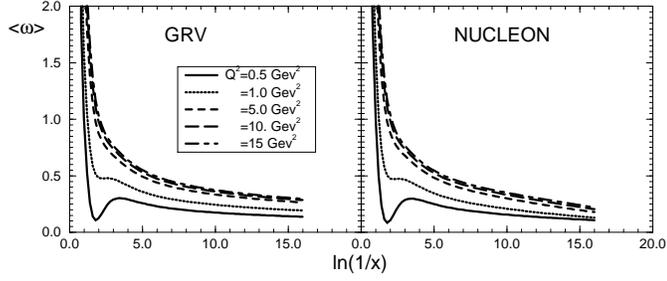,width=90mm,clip=}}
\caption{\em The effective power $<\omega >$
 calculated for $xG^A(A = 1)$ and the GRV 
distribution. }
\label{omn}
\end{figure}

\begin{figure}[htbp]
\centerline{\epsfig{figure=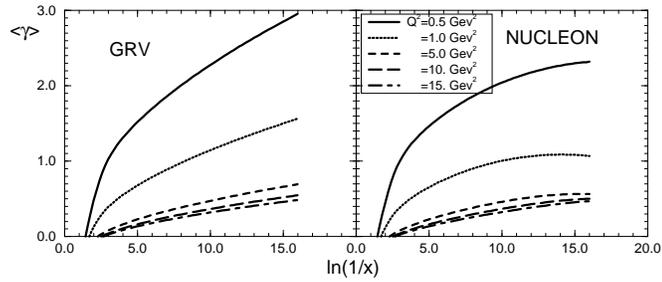,width=90mm}}
\caption{\em The effective power of $Q^2$ dependence calculated for
 $xG^A(A = 1)$ 
and the GRV  distribution. }
\label{gmn}
\end{figure}

The eq.(\ref{omega}) gives the average value of the effective power 
$<\omega > $  of the  gluon distribution, 
$xG(x,Q^2) \propto x^{-<\omega>} $, which is
suitable to study the small $x$ behavior of the gluon distributions. 
Fig.\ref{omn} shows the calculation of $<\omega>$ the nucleon distribution
for  \eq{FINANS}
and for  GRV gluon distribution, both as  functions of $ln(1/x)$ for different 
values of $Q^2$. From the figure, we can see that the effective powers 
of $xG^A(A=1)$ and $xG(GRV)$ have the same general behavior in the small $x$ 
limit but the nucleon distribution is slightly suppressed.
We calculate also, in the same kinematical region, the exponent $<\gamma>$, 
given by eq (\ref{gama}). This  is the average value of the anomalous 
dimension, which describes the effective dependence of the distribution in 
$Q^2$ variable. Figs.\ref{gmn} shows $<\gamma>$ for  the 
nucleon and GRV distributions, indicating that
 the $Q^2$ dependence is slightly soften by the SC.

Comparing figures \ref{omn} and \ref{gmn}, we can conclude that even these 
more detailed characteristic of the gluon structure function have not been
seriously affected by the SC in the nucleon case. 

\begin{figure}[htbp]
\centerline{\epsfig{figure=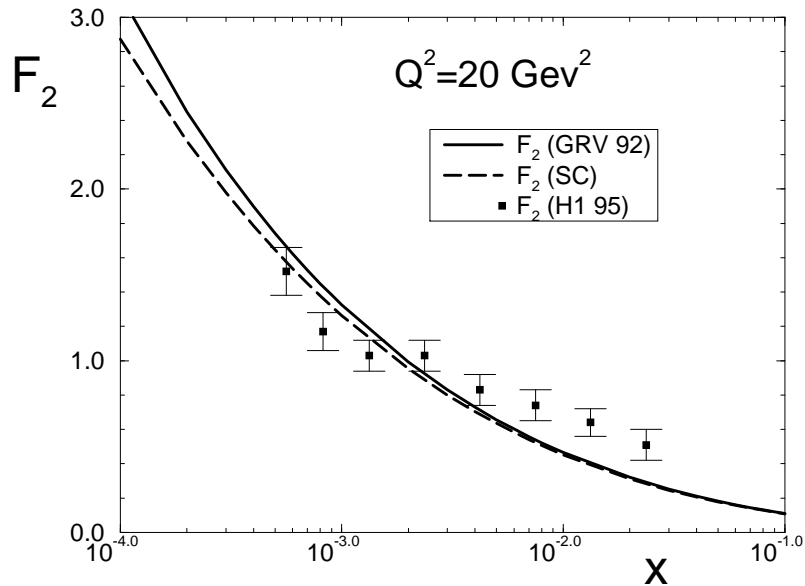,height=90mm}}
\caption{\em $F_2$  from $xG^A$  and  the  GRV 
distribution, compared with experimental data \protect\cite{HERA}. }
\label{f21}
\end{figure}

We also use the DGLAP evolution equations to predict the value of the deep
inelastic structure function $F_2$ from the $xG^A$ gluon distribution.
 Summing the DGLAP
evolution equations for each quark flavor, 
the function $F_2$ may be written \cite{r30}
\bea
F_2 =\frac{\as (Q^2)}{\pi} \sum_{q} e_{q}^{2} \int^{{Q}^2}_{{Q}^{2}_{0}}
 \frac{d Q'^2}{Q'^2} 
\int^{1-x}_{0} [ z^2 + (1-z)^2 ] \frac{x}{1-z} G^{N}(\frac{x}{1-z}, {Q'}^2)
\label{eqf21}
\eea
where the sea quark distributions   have been  neglected in comparison with 
the gluon distribution.
Fig.\ref{f21} shows the prediction for $F_2$ from $xG^A$  and from the  GRV 
distribution, compared with experimental data. As we can see, the magnitude of 
the suppression due to the  SC is less than $10 \%$ in the region of the 
HERA data and this suppression 
is smaller than the experimental error.

From the above results we can conclude that 
\eq{FINANS} gives a good description for the gluon structure 
function for nucleon and describes the available experimental data.
 Therefore, it can be taken as 
a correct first approximation
in the approach to the nucleus case.

{\em 3.6 The gluon structure function for nucleus.}

In the framework of perturbative approach
it is only possible to calculate 
the behavior of the gluon distribution at small distances. The initial gluon 
distribution should be taken from the experiment. Actually the initial
virtuality $Q_0^2$ should be big enough to guarantee that we are dealing
with the leading twist contribution. Our main assumption is that we start
the QCD evolution with a small value of $Q_0^2$ considering that the 
 MF is a good
model for high twist contributions in   DIS off nucleus.

The scale of the SC governs by the value of $\kappa_A$, namely
they are big for $\kappa_A\,>\,$1 and  small for $\kappa_A\,<\,$ 1.
Fig.4 shows the plot of $\kappa_A$ = 1 for different nuclei. One can see
that the SC should be essential for heavy nuclei starting from Ca at the
accessible experimentally kinematic region.

\begin{figure}[p]
\centerline{\epsfig{figure=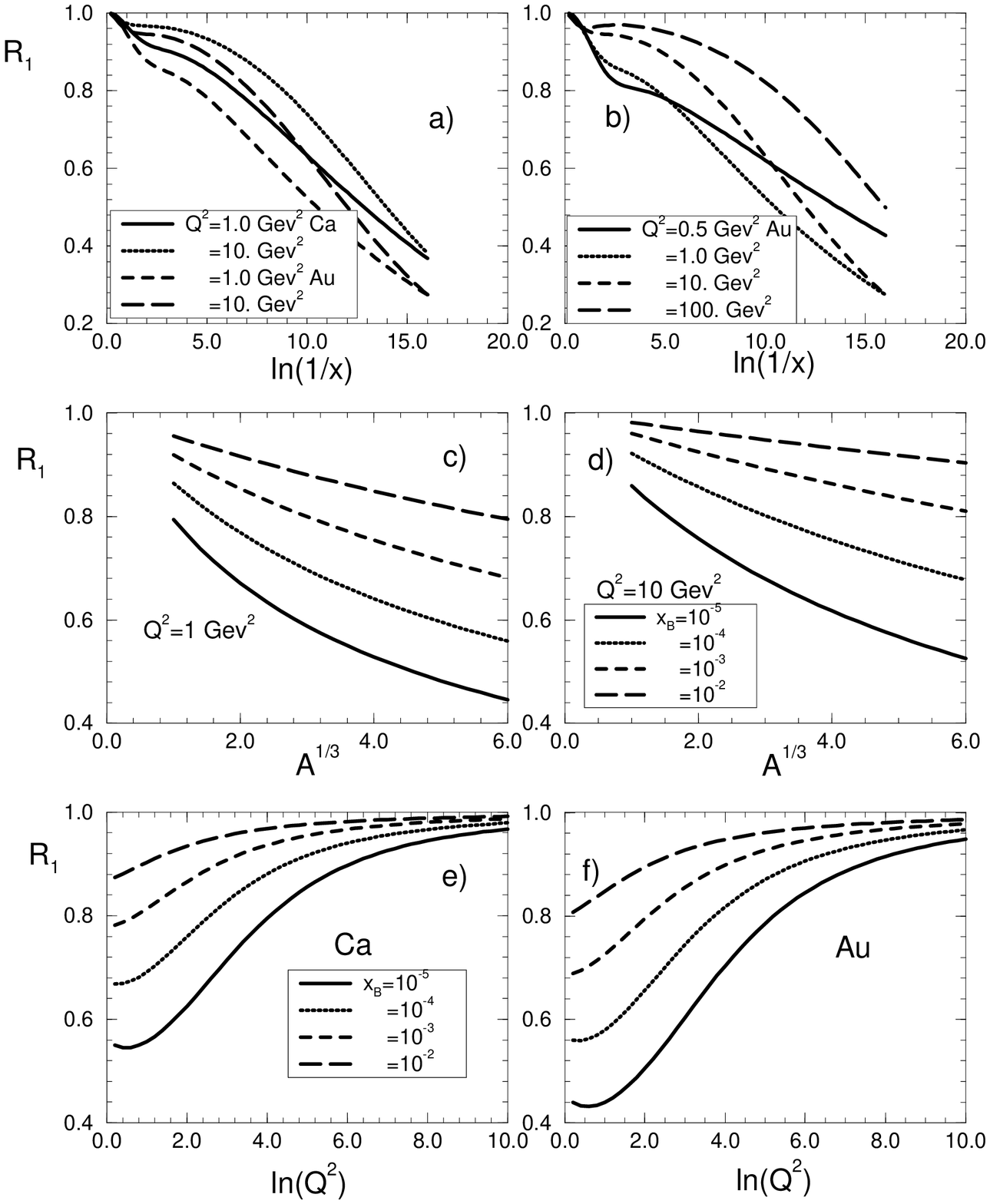,width=150mm,height=90mm}}
\caption{\em $R_1$ as a function of $ln(1/x)$,
 $lnQ^2$ and $A^{1/3}$: a) $R_1$ as a function of $ln(1/x)$ 
for different nucleus and different values of A; b)
$R_1$ as a function of
$ln Q^2$ for different values of $x_B$ for Au;
 c)  and  d) $R_1$ as a function of $A^{1/3}$ for different $Q^2$; e) and f)
 $R_1$ dependence on $Q^2$ for Ca and Au.}
\label{r1a}
\end{figure}

Now we extend the definition of $R_1$ for the nucleus case
\beq
R_1\,\,=\,\,\frac{xG^A(x,Q^2)}{A x G^{GRV}_N (x,Q^2)}\,\,,
\label{ratr1}
\eeq
where the numerator is calculated using eq.(\ref{FINANS}). Figure \ref{r1a}
shows the results for the calculations of $R_1$ as a function of
the variables $ln(1/x)$, $lnQ^2$ and $A^{1/3}$. Fig.\ref{r1a}a presents the 
ratio $R_1$ for two different values of $Q^2$ and for different nuclei.
The suppression due to the SC increases with $ln(1/x)$ and is much bigger than
 for
the nucleon case. For $A=40$ (Ca) and $Q^2=10 \, GeV^2$, the suppression varies
from $4 \, \%$ for $ln(1/x)= 3$ to $25 \, \%$ for $ln(1/x)= 10$. For $A=197$ 
(Au) the suppression is still bigger, going from $6 \%$ to $35 \%$ in the same
kinematic region. Fig. \ref{r1a}b shows the same ratio for different values of 
$Q^2$ for the gold. The suppression decreases with $Q^2$.
Figs. \ref{r1a}c and d show  the $R_1$
ratio as a function of $A^{1/3}$ and $x$ for a fixed value of $Q^2$.
As 
expected, the SC increases with $A$. An interesting feature of this figure 
is the
fact that the curves tend to straight lines as $x$ increases. It occurs 
because, as $x$ grows, the structure function $xG (GRV)$ becomes smaller, 
and the correction term of (\ref{FINANS}) proportional to $\kappa$ dominates. 
Since $\kappa$
is proportional to $A^{1/3}$, the curves behave as straight lines.
The decrease of suppression with $Q^2$ is illustrated in more detail in Figs.
\ref{r1a}e and f which presents $R_1$ as a function of $\ln Q^2$ for different
values of $x$ for Ca and Au, respectively. 
The effect is pronounced for small $Q^2$ and 
$x$ and diminishes as $ln Q^2$ increases.

\begin{table}
{\bf Table 1:} Values of $R_{1 N}$ and $\alpha$ for parameterization 
$R_1 \,=\,R_{1N}\,A^{-\,\alpha}$.\\
\vspace{0.5cm}
\begin{center}
\begin{tabular} {||l|l|l|l|l||}
\hline
\hline
\multicolumn{1}{||c|}{} & \multicolumn{2}{c|}{$Q^2=1 GeV^2$}
 & \multicolumn{2}{c||} {$Q^2 = 10 GeV^2$}
\\
\cline{2-5}
$x$ & $R_{1N}$ & $\alpha$ & $R_{1N}$ & $\alpha$ \\
\hline
$10^{-2}$ & 0.94   & 0.0416 & 0.98 & 0.014\\
\hline
$10^{-3}$ & 0.92 & 0.0616 & 0.94 & 0.034\\
\hline
$10^{-4}$ & 0.88 & 0.094 & 0.92 & 0.0563\\
\hline
$10^{-5}$ & 0.8 & 0.145 & 0.86 & 0.093\\ 
\hline
\hline
\end{tabular}
\end{center}
\end{table}
This picture ( Fig.\ref{r1a} ) shows also that the gluon structure function is
far away from the asymptotic one. The asymptotic behavior $R_1\,\rightarrow 1$
 ( see Figs.\ref{r1a}e and f )
occurs only at very high value of $Q^2$ as well as in the GLR approach
( see ref. \cite{I3} ). The asymptotic $A$-dependence (  $R_1\,\propto\,
A^{-\frac{1}{3}}) $ ) has not  been seen in the accessible kinematic range of
 $Q^2$ and $x$ ( see Figs. \ref{r1a}c and d and Table 1 ). This result also
 has been predicted in the GLR approach \cite{I2}. We want also to mention that
parameterization $R_1\,=\,R_{1N}\,A^{- \alpha}$ does not fit the result
 of calculations   quite well for
$1Gev^2\,\leq \,Q^2\,\leq\, 20 GeV^2$ and $ 10^{-2}\,\leq\,x\,\leq\,10^{-5}$.
For $x\,\sim\,10^{-2}$ the parameterization $R_{1}\,=\,R_{1N}
 \,-\,R' \,A^{\frac{1}{3}}$ with 
parameters $R_{1N}$
 and $R'$ for each value of $Q^2$,  works much better reflecting that
 only the first correction to the Born term is essential in the Mueller formula.

\begin{figure}[hptb]
\centerline{\epsfig{figure=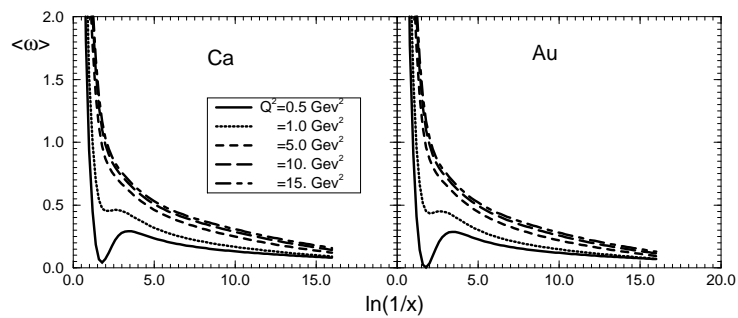,width=100mm}}
\caption{\em $<\omega>$ for different values of $Q^2$ and $A$. }
\label{oma}
\end{figure}

We extend also the calculation of the exponents $<\omega>$ and $<\gamma>$
of the semiclassical approach
for the nuclear case. We calculate the effective power of the nuclear gluon
distribution $<\omega>$ using the expression
\beq
< \omega >\,\,=\,\,\frac{\partial \ln (x G^A (x,Q^2))}{\partial \ln(1/x)}\,\,.
\label{omegaa}
\eeq
Fig.\ref{oma} shows the results as functions of $ln(1/x)$ for different values 
of $Q^2$ and different nucleus. The SC decreases the effective power of
 the nuclear distribution, giving rise to a flattening of 
the distribution in the 
small $x$ region.

 It is also interesting to notice that at small values of $Q^2$, the
effective power tends to be rather small, even in the nucleon case, at
very small $x$. However it should be stressed that the effective power
remains bigger than the intercept of the so called "soft" Pomeron \cite{SOFTPO},
even in the case of a sufficiently heavy nucleus (Au), for $Q^2\,>\,1GeV^2$.
Nowadays, many parameterizations \cite{CAPELA} with matching of "soft" and
"hard" Pomeron have appeared triggered by new HERA data on diffraction
dissociation \cite{DDHERA}. These parameterization used Pomeron-like behavior
namely, $xG(x,Q^2) \propto x^{- \omega(Q^2)}$. However, if the Pomeron is a 
Regge pole, $\omega$ cannot depend on $Q^2$, and the only reasonable
explanation is to describe $\omega(Q^2)$ as the result of the SC. Looking at
Fig.\ref{oma} we can claim the SC from the MF cannot provide sufficiently 
strong SC  to reduce the value of $\omega$ to $0.08$, a typical value 
for the soft Pomeron \cite{SOFTPO}, at least for $Q^2 \geq 1 GeV^2$.

The calculation of the effective value of the anomalous
dimension $\gamma$ may help us to estimate  what distances work in the
SC corrections. This effective exponent is given by
\beq 
< \gamma >\,\,=
\,\,\frac{\partial \ln (x G^A(x,Q^2))}{\partial \ln (Q^2/Q^2_0)}\,\, .
\label{gma1}
\eeq 
\begin{figure}[htbp]
\centerline{\epsfig{figure=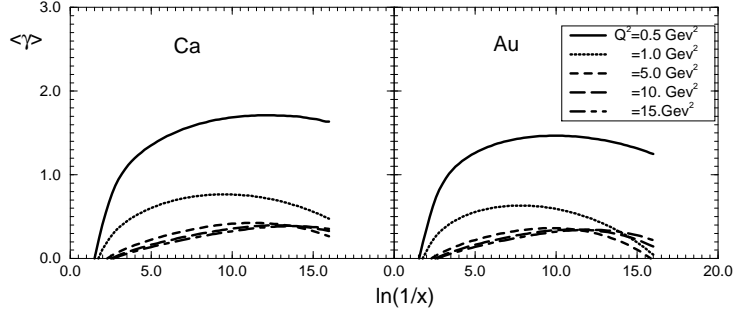,width=100mm}}
\caption{\em $< \gamma > $ for different $Q^2$ and $A$. }
\label{gma}
\end{figure}

Fig.\ref{gma} shows the results as functions of $ln(1/x)$ for different
values of $Q^2$ and for two nuclei.  We see that the values of $\gamma$ at 
$ln(1/x)\, \leq \,5$, for both Ca and Au, is very close to the results for 
GRV  and for nucleon case.  At smaller values of $x$, the anomalous
dimension presents a sizeable reduction, which increases with A. For
$ln(1/x)\, > \,15$, $<\gamma>$ tends to zero unlike in the DGLAP evolution
equations ( see Fig.10 for the GRV parameterization). Analysing the $Q^2$ 
dependence, we see that $<\gamma>$ is bigger than $1$ only for 
$Q^2= 0.5 \, GeV^2$. For $Q^2= 1.0 \, GeV^2$, the anomalous
dimension is close to $1/2$, and for  $Q^2 > 5.0 \, GeV^2$ it is always
smaller than $1/2$. 

 Using semiclassical approach, we see that

\beq
\kappa \propto \frac{1}{Q^2}( Q^2)^{\g}\, ,
\label{kappa}
\eeq
and if $\gamma \geq 1$, the integral over $r_t$ in  the master equation
(\ref{FINANS}) becomes divergent, concentrating at small distances.

If  $ 1\,>\,\gamma \geq 1/2$, only the first SC term, namely,
 the second term in expansion
of the master equation, is concentrated at small distances, while higher
order SC are  still sensitive to small $r_{\perp}$ behavior. 
Fig.\ref{gma} shows
that this situation occurs for $Q^2 \, > \, 1 \, GeV^2$, and even for 
$Q^2 = 1 GeV^2$ at very small values of $x$. 
We will return to discussion of these properties of the anomalous 
dimension behavior in the next section.

{\em 3.7 The gluon life time cutoff.}

In the DIS  the
incident electron penetrates the nucleus and radiates the virtual photon
 whose lifetime $\tau_{\gamma^*}\,\propto\,\frac{1}{ m x}$ \cite{r24}.
We can recover three different kinematic regions:

1. $\tau_{\gamma^*}\,=\,\frac{1}{ m x}\,\,<\,\,R_{NN}$, where $R_{NN}$ is
the characteristic distances between the nucleons of the nucleus.
This virtual photon can be absorbed only by one nucleon and the total cross 
section is  $\s(\gamma^* A) \,=\,A \,\s( \gamma^* p)$.

2. $ R_A\,\,>\,\,\tau_{\gamma^*}\,=\,\frac{1}{ m x}\,\,>\,\,R_{NN}$, where
 $R_A$ is the nucleus radius.
In this kinematic region the virtual photon can interact with the group of
nucleons. However, $\s(\gamma^* A)$ is still proportional to $A$ since the
number of nucleons in a group is much less than $A$.

3. $\tau_{\gamma^*}\,=\,\frac{1}{ m x}\,\,>\,\,R_{A}$. Here, before reaching
the front surface of the nucleus, the virtual photon ``decomposes"
 in the developed parton cascade which then interacts with the nucleus.
It can be shown \cite{I1} that the absorption cross section of the virtual
 photon will now be proportional to the surface area of the nucleus
$\s(\gamma^* A)\,
\propto\, A^{\frac{2}{3}} $, because the wee partons of the parton cascade
are absorbed at the  surface and do not penetrate into the centre of
 the nucleus.

Everything that we have discussed have been calculated in the third 
kinematic region. For the RHIC energies we have to develop some technique how 
to penetrate into the second one. To do this we have to remember that the
 opacity 
 ( or $\kappa_G$ ) actually depends on the longitudinal part of the momentum transfer
( $q_z$ ) which could be calculated in terms of $x$ and $x'$ of our master
 equation (\ref{MF}) , namely, $q_z\,=\, (x + x') m $ ( see Ref.\cite{AGL} ).
 Recalling that opacity $\Omega \,\propto\,r^2_{\perp} x G( x,q_z, 
\frac{1}{r^2_{\perp}})\,\,S(b_t,q_z)$ we see that $q_z$ - dependence enters
to two factors: to gluon structure function and to the nucleon profile 
function. We know how to take into account the $q_z$ - dependence of the
gluon structure function ( see Ref. \cite{GLR} where the DGLAP equation
for $q_z\,\neq\,0$ is written ). However we neglected this effect in our
 present estimates, hoping that this dependence occurs on the hadron scale 
and cannot change too much the dependence of the SC on the number of collisions 
during the passage through the nucleus. 

 The dependence of the profile function $S(b_t,q_z)$ on $q_z$
have been discussed and
in the Gaussian parametrization
it can be factor out in the form:
\beq \label{QZFORM}
S(b_t,q_z)\,\,=\,\,S(b_t)\,\cdot\,L(q_z)\,\,=\,\,S(b_t)\,\cdot\,
e^{-\,\frac{R^2_A}{4}\,q^2_z}\,\,.
\eeq 
This $q_z$ - dependence takes into account the fact that
 the virtual gluon  can interact with
 the target only during  the finite time $\tau \,=\,1/mx$ undergoing
$\rho \,\tau \,<\,\rho\,R_A$ collisions.  Using \eq{QZFORM}, we can obtain:
\beq \label{cut}
x \,G_A( x, Q^2)\,\,=\,\, A\,x\,G_N(x, Q^2) \,\,-\,\,A\,\frac{\as\,N_c}{\pi}
\int^1_x\,\,\int\,\frac{d x'}{x'}\,\frac{d Q'^2}{Q'^2}\,\,
L(q_z)\,\,x'\,G_N(x',Q'^2)
\eeq
$$
+\,\,
\frac{ 2 R^2_A}{\pi^2}\,\int^1_x\,\frac{ d x'}{x'}\,\int^{\frac{1}{Q^2_0}}_{
\frac{1}{Q^2}}\,\frac{d r^2_t}{r^4_t}\,\,\{\,C\,+\,
\ln(L(q_z)\,\kappa_G(x',r^2_t))
+\,\,E_1(L(q_z)\, \kappa_G(x',r^2_t)\,)\}
\,\,.
$$

\begin{figure}[hptb]
\centerline{\epsfig{figure=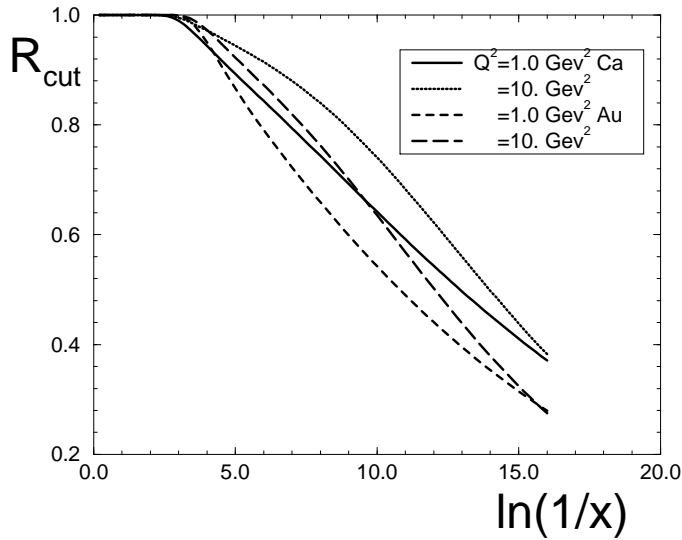,width=100mm}}
\caption{\em $ R_1$  for CA and Au with gluon life time cutoff.}
\label{r1cut}
\end{figure}

Fig. \ref{r1cut} shows the result of our calculations.
Comparing Fig.\ref{r1a} with this picture, one can see that the finite life
 time of the virtual gluon affects  the behavior of the gluon structure
function only at sufficiently large $x$ ( $x\,\geq\,10^{-2}$ ) diminishing
the value of the SC in this kinematic region. 
 This effect turns out to be very important for the RHIC energies and
has to be studied in more details. 

\section{ First corrections to  the Glauber  ( Mueller) Approach.}
\label{cgf}
In this section we discuss  the corrections to the Glauber approach
 (the  Mueller formula of \eq{MF} .
To understand how big could be the corrections to the Glauber approach
 we calculate the second iteration of the Mueller formula of \eq{MF}. As
has been discussed,\eq{MF} describes the rescattering of the fastest gluon
( gluon - gluon pair ) during the passage through a nucleus
 ( see Fig.1 ). In the second iteration we take into account also
the rescattering of the next to the fastest gluon. This is a well defined task
 due to the strong ordering in the parton fractions of energy in the
parton cascade in leading $ln (1/x)$ approximation of pQCD that we are 
dealing with. Namely:
\beq \label{95}
x_B\,\,<\,\,x_n\,\,<\,\,...\,\,<x_1\,\,<\,\,1\,\,;
\eeq
where 1 corresponds to the fastest parton in the cascade.

\begin{figure}[htbp]
\centerline{\epsfig{figure=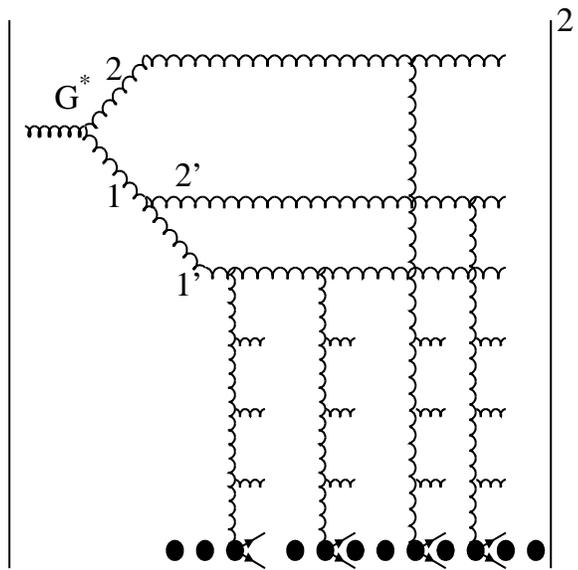,height=100mm}}
\caption{\em  The interaction with nucleons that is  taken into
account in the second iteration of 
M ueller formula. }
\label{Fig.5}
\end{figure}

Therefore, in the second interaction we include the rescatterings of the 
gluons with the energy fraction $1$ and $x_1$ ( see Fig.\ref{Fig.5} ).
 Doing the first
iteration we insert in \eq{MF} $G_N(x,Q^2)\,=\,G^{GRV}_{N}(x,Q^2)$. For the
second iteration we calculate the gluon structure function using \eq{MF}
 substituting
\beq \label{96}
x \,G_N\,\,=
\,\,\frac{x G^1_{A} (x, Q^2)}{A}\,\,-\,\,x\, G^{GRV}_{N} ( x,Q^2)\,\,;
\eeq
where $xG^1_A$ is the result of the first iteration of \eq{MF} that has been
discussed in details in section 3.

\begin{figure}[hptb]
\centerline{\epsfig{figure=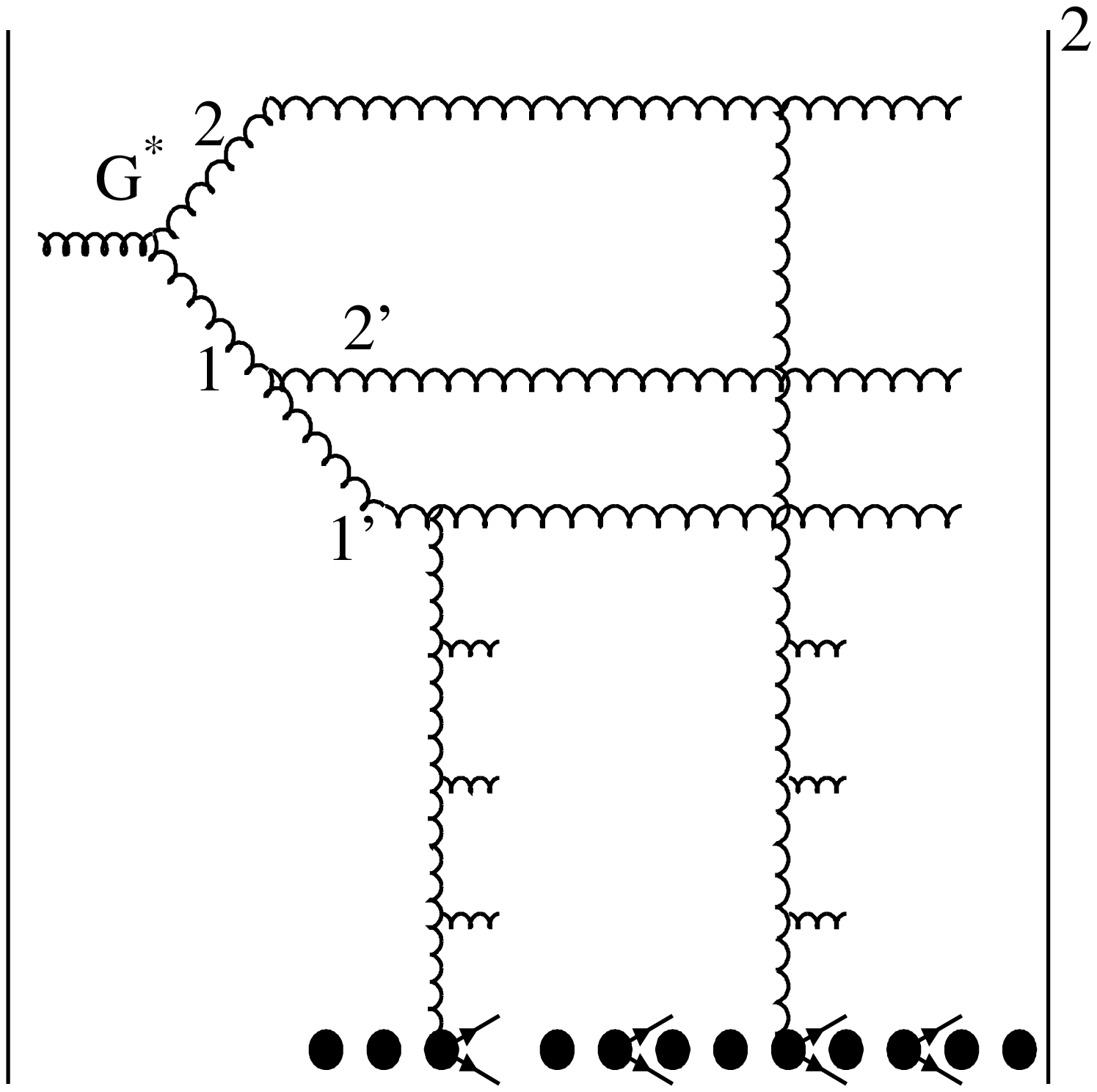,width=100mm}}
\caption{ \em The first term of the  second iteration 
 of \protect\eq{96}.}
\label{fig.16}
\end{figure}

Fig.\ref{fig.16}
  shows the need to subtract  $ xG^{GRV}_N$ in \eq{96} making the
second iteration. Indeed, in the second iteration we take into account the
rescattering of gluon $1'$ - gluon $2'$ pair off a nucleus. We picture
 in Fig.\ref{fig.16}  the
 first term of such iteration in which $G_{1'} G_{2'} $ pair has no
 rescatterings. It is obvious that it has been taken
into account in our first iteration, so we have to subtract it to avoid a
 double counting.

\begin{figure}[hptb]
\begin{center}
\begin{tabular}{ c c}
\epsfig{file=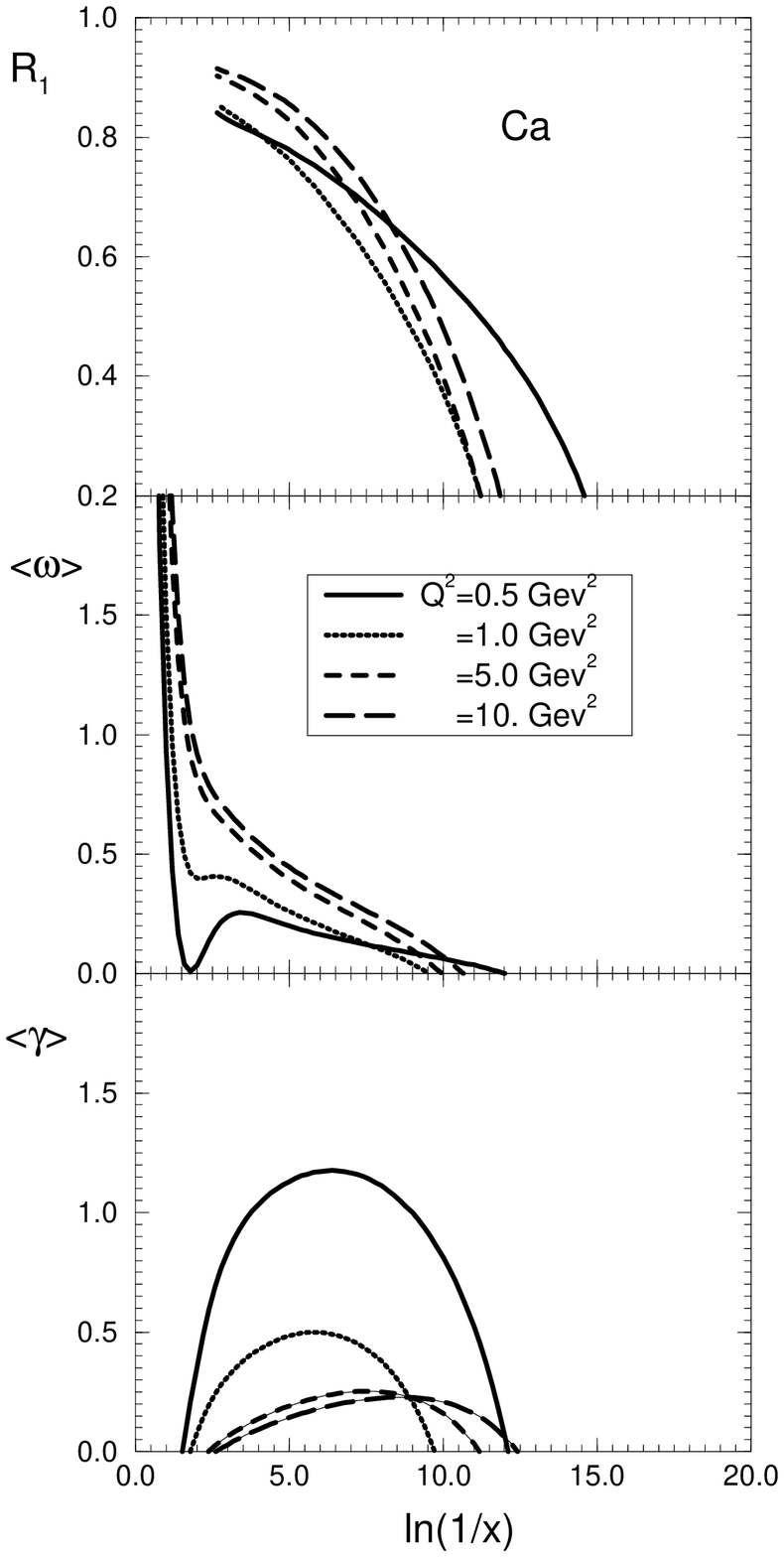,width=80mm} & 
\epsfig{file=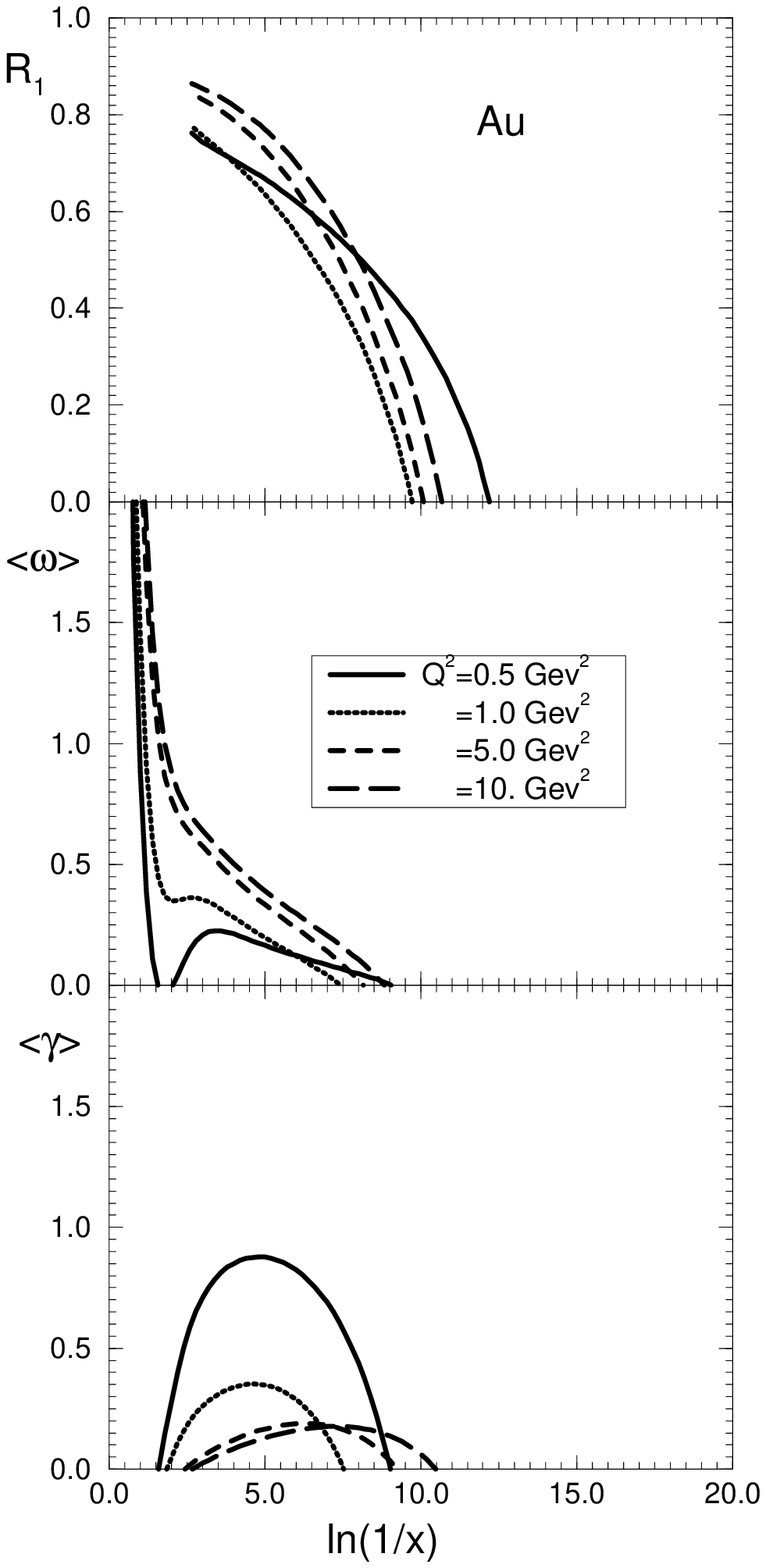,width=80mm}\\
\end{tabular}
\end{center}
\caption{ \em Second iteration calculations for $R_1$, $<\o>$, and $<\gamma >$
 for Ca and Au.}
\label{2ndit}
\end{figure}

One can see in Figs.\ref{2ndit} that the second iteration gives a big effect 
and changes crucially $R_1$, $< \gamma >$, and $<\omega >$.
The most remarkable feature
 is the crucial change of the value of the effective power $\omega(Q^2)
$ for the ``Pomeron" intercept which tends to zero at HERA kinematic region,
making possible the matching with ``soft" high energy phenomenology.
It is also very instructive to see how the second iteration makes
more pronounced all properties of the behavior of the anomalous dimension ( $<
\gamma>$) that we have discussed.
The main conclusions which we can make from Figs.\ref{2ndit} are:
(i) the second iteration gives a sizable contribution in the region $x\,\,<\,
\,10^{- 2}$ and for $x\,\,\leq \,\,10^{-3} $ it becomes of the order of the
first iteration; (ii) for $x\,\,<\,\,10^{-3}$ we have to calculate the
 next iteration. It means  that for such small $x$ we have to develop a 
different technique  to take into account rescatterings of  all the partons
 in the parton cascade which will be more efficient than the simple iteration
procedure for \eq{MF}. However, let us first understand why the second iteration
becomes essential to establish small parameters that enter to our problem.

\begin{figure}[p]
\begin{center}
\begin{tabular}{c c}
\epsfig{file=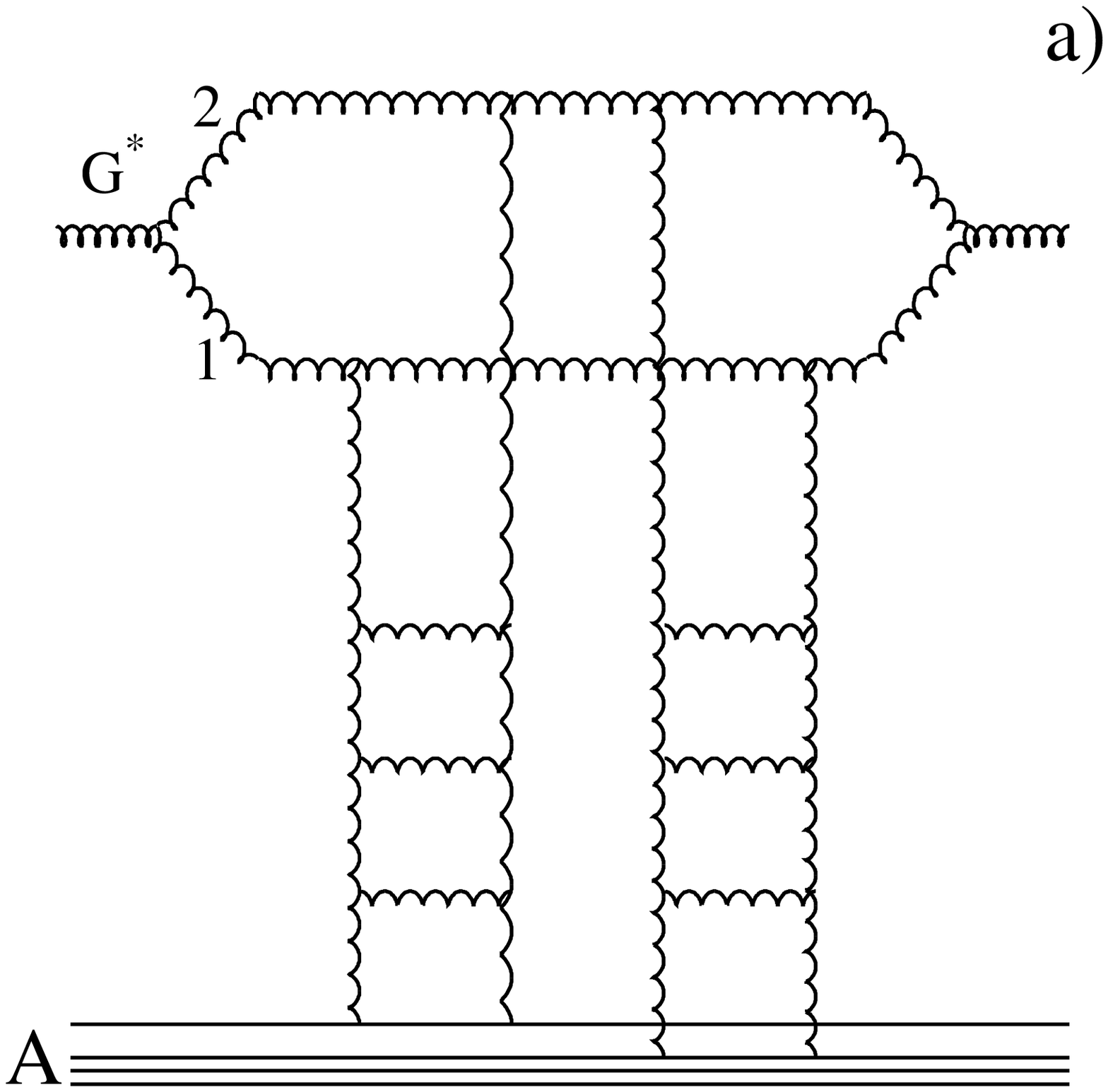,width=6.5cm} &
\epsfig{file=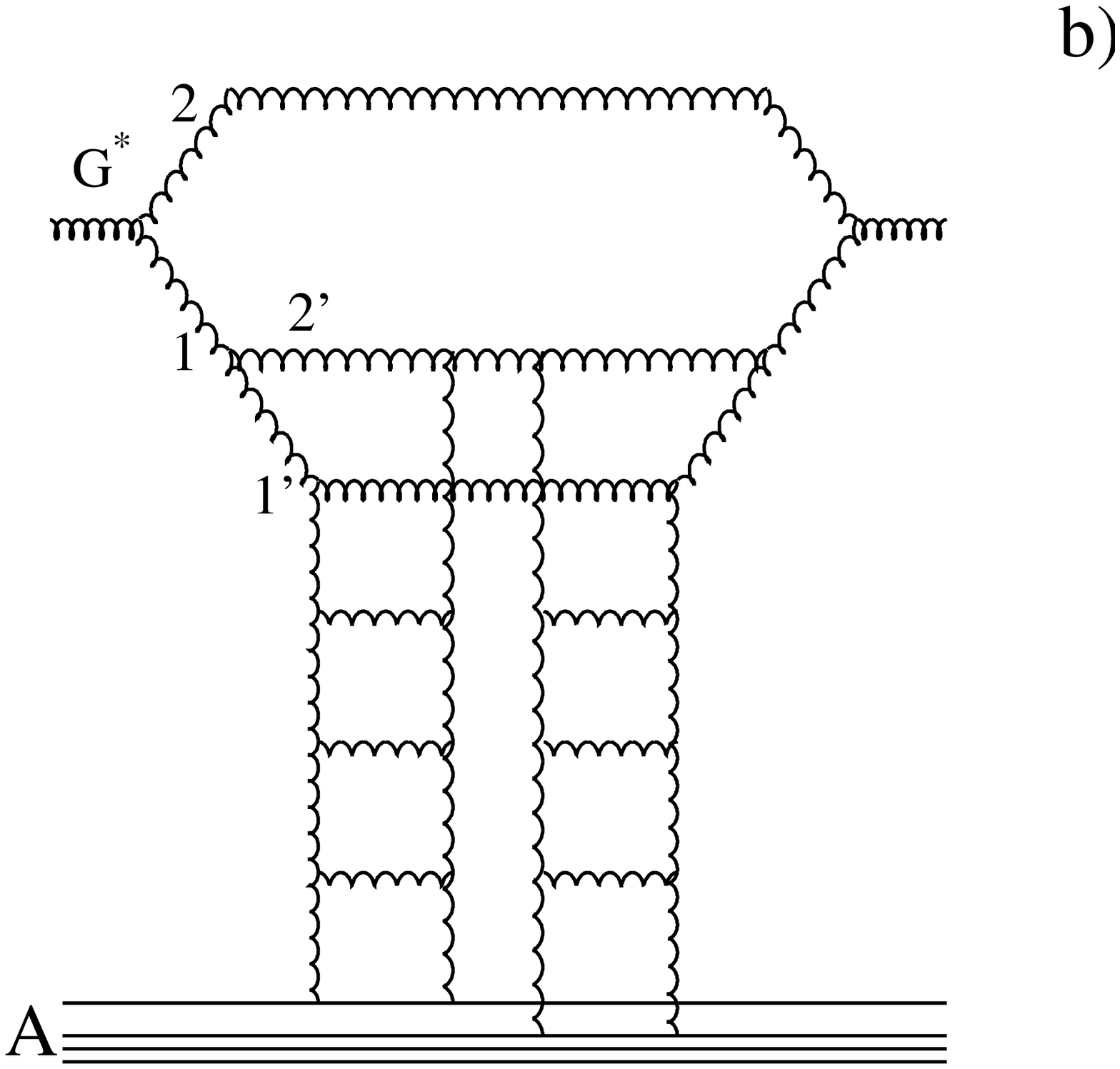,width=6.5cm}\\ 
\epsfig{file=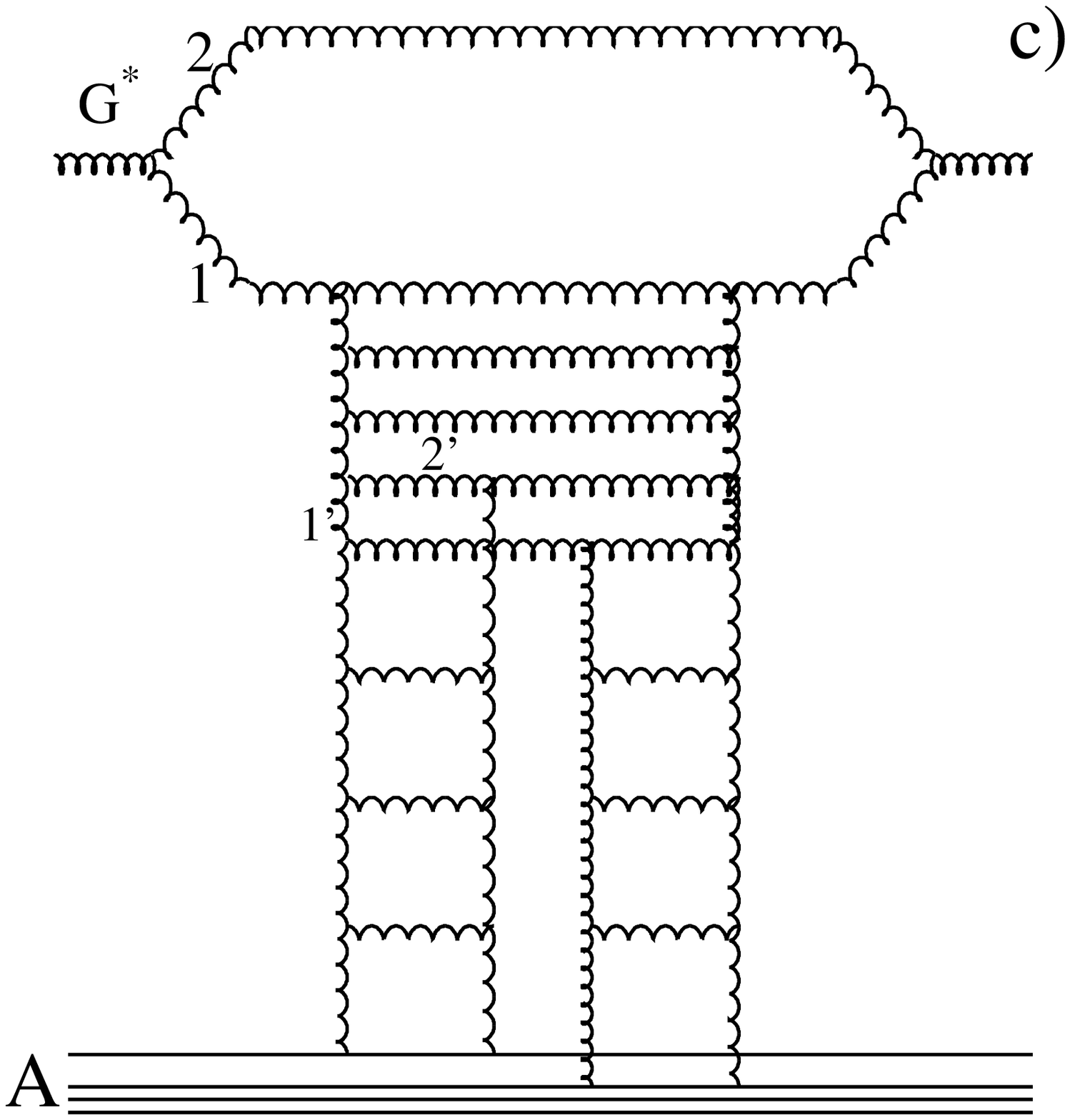,width=6.5cm} &
\epsfig{file=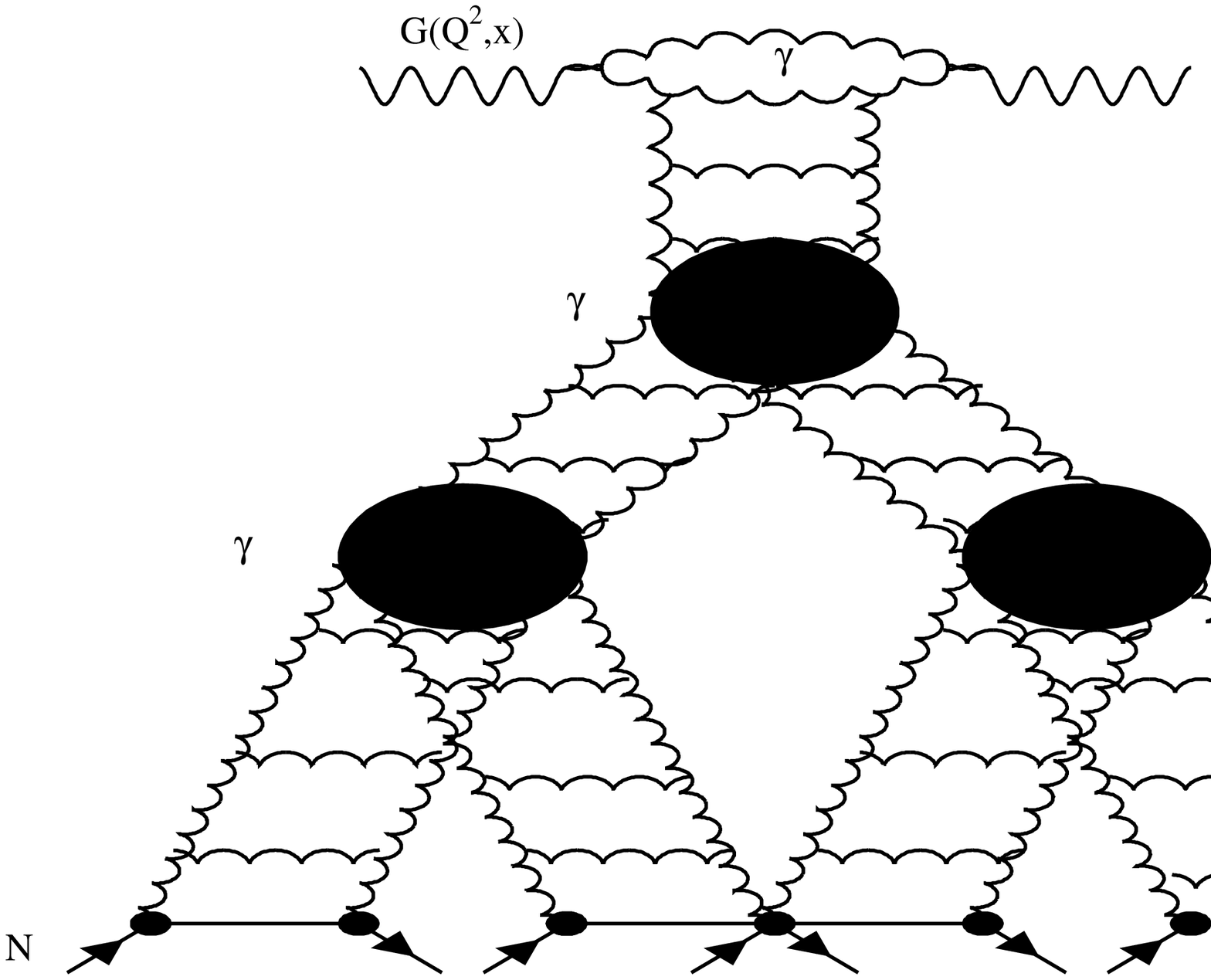,width=6.5cm}\\ 
\end{tabular}
\end{center}
\caption{\em Corrections to the Glauber approach.}
\label{fig.18}
\end{figure}

As has been discussed, we use the GLAP evolution equations for gluon 
structure function in the region of small $x$. It means, that we sum the Feynman
diagrams in pQCD using the following set of parameters:
\beq \label{97}
\as\,\,\ll\,\,1\,\,;\,\,\,\,\,\,\,\,\,\,\as\,\ln\frac{1}{x}\,\,<\,\,1\,\,;
\,\,\,\,\,\,\,\,\,\,\as\,\ln\frac{Q^2}{Q^2_0}\,\,<\,\,1\,\,;
\,\,\,\,\,\,\,\,\,\,
\as\,\,\ln\frac{Q^2}{Q^2_0}\,\,\ln\frac{1}{x}\,\,\approx\,\,1\,\,.
\eeq
The idea of the theoretical approach of rescattering that
 has been formulated in
 the GLR paper \cite{GLR} is to introduce a new parameter \footnote{
In the GLR paper the notation for $\kappa$ was W, but in this paper 
we use $\kappa$ to avoid a misunderstanding since, in DIS, W is the energy
of interaction.}:
\beq \label{98}
\kappa\,\,=\,\,\frac{N_c \,\as\,\pi \,A}{ 2\,Q^2\,R^2_A}\,x G (x,Q^2)
\eeq
and sum all Feynman diagrams using the set of \eq{97} and $\kappa$ as parameters
of the problem, neglecting all contributions of the order of: $\as$,
 $\as \,\kappa$,
$\as \ln(1/x)$, $\as \ln(1/x)\,\kappa$, $\as \ln (Q^2/Q^2_0)$ and 
 $\as \ln (Q^2/Q^2_0)\,\kappa$. It should be stressed that Mueller formula
gives a solution  for such  approach. Indeed, \eq{MF} depends only on
$\kappa$ absorbing all \\
$(\,\as\,\ln (Q^2/Q^2_0)\,\ln(1/x)\,)^n$
contributions in $x G(x,Q^2)$. However, it is not a complete solution.
To illustrate this point let us compare the value of the second term of the
expansion of \eq{MF}  with respect to $\sigma (r^2_t) $ with the first
correction due to the second iteration in the first term of such an expansion.
In other words we wish to compare the values of the diagrams in Fig.\ref{fig.18}
b and Fig.\ref{fig.18}a.
The contribution of the diagram of Fig.\ref{fig.18}a is equal:
\beq \label{99}
\Delta x G(x, Q^2)\,(\, Fig. \ref{fig.18}a
\, ) \,\,=\,\,\frac{R^2_A}{2\,\pi^2 }\,\,
\int \frac{d x'}{x'}\,\int\,d Q'^2 \,\,\kappa^2 ( x', \frac{Q'^2}{4})\,\,,
\eeq
where $x'$ and $Q'^2$ are the fraction of energy and the virtuality of 
 gluon $1$ in Fig.\ref{fig.18}a

The diagram of Fig.\ref{fig.18}b
 contains one more gluon and its contribution is:
$$
\Delta x G(x, Q^2)\,(\, Fig.\ref{fig.18}b
\, ) \,\,=\,\,\frac{R^2_A}{\pi^2 2}\,\,
\frac{N_c \as}{\pi} \int\,\,\frac{d x'}{x'} \,\,\frac{ d Q'^2}{Q'^2}\,\,
\int \frac{d x"}{x"}\,\int\,d Q"^2 \,\,\kappa^2 (x",\frac{ Q"^2}{4})
$$
\beq \label{100}
\propto\,\,\frac{\as N_c}{\pi}\,\,\ln(1/x)\,\,\ln(Q^2/Q^2_0)\,
\,\Delta x G(x,Q^2)
(\,Fig.\ref{fig.18}a)\,)\,\,,
\eeq
where $x'$ ($x''$) and $Q'^2$ ( $Q''^2$) are the fraction of energy and the
virtuality of gluon  $1$ ($1'$) respectively in Fig.\ref{fig.18}b.
Therefore, \eq{100} gives the contribution which is of the order of
\eq{99} in the kinematic region where the set of parameters of \eq{97} holds.
It means also that we need to sum all diagrams of Fig.\ref{fig.18}b
 type to obtain 
the full answer. In the diagram of Fig.{\ref{fig.18}b
 not only one but many gluons
 can be emitted. Such emission leads to so called ``triple ladder" interaction,
pictured in Fig.\ref{fig.18}c ( see ref.\cite{GLR} ). This diagram is the first
from so called ``fan" diagrams of Fig. \ref{fig.18}d. 
To sum them all we can neglect
 the third term in \eq{MF} and treat the remained terms as an equation
 for $x G(x,Q^2)$. It is easy to recognize that we obtain the GLR equation
  \cite{GLR}\cite{MUQI}. Generally speaking the GLR equation sums the most
important diagrams in the kinematic region where $\as\,\ln(1/x)\,\ln(Q^2/Q^2_0)
\,\,\gg\,\,1$ and $\kappa\,<\,1$.

\section{The general approach.}

{\em 5.1 Why equation?}

We would like to suggest a new approach based on the new evolution equation
 to sum all SC. However, first of all we want to argue why an equation is
 better than any iteration procedure. To illustrate this point of view let us
differentiate the Mueller formula with respect to $y\,=\,\ln(1/x)$ and $
\xi\,=\,\ln Q^2$. It is easy to see that this derivative is equal to
\beq \label{DER}
\frac{\partial^2 x G(x, Q^2)}{\partial y\, \partial \xi}\,\,=\,\,
\frac{4}{\pi^2}\,\int d b^2_t \,\,\{\,\,1\,\,-\,\,e^{-\,\frac{1}{2}\,\s(x,
r^2_{\perp}\,=\,\frac{1}{Q^2})\,S(b^2_t)}\,\,\}\,\,.
\eeq
The nice property of \eq{DER} is that everything enters at small distances,
therefore everything is under theoretical control. Of course, we cannot get rid
of our problems changing the procedure of solution. Indeed, the nonperturbative
effects coming from the large distances are still important but they are all
 hidden in the boundary and initial conditions to the equation. Therefore,
an equation is a good ( correct ) way to separate  what we know ( small
distance contribution) from what we don't ( large distance contribution).

{\em  5.2 The generalized evolution equation.}

We suggest the following way to take into account the interaction of all
partons in a parton cascade with the target. Let us differentiate the Mueller
formula over $y \, = \,\ln (1/x)$ and $ \xi = \ln(Q^2/Q^2_0)$. It gives:
\beq \label{102}
\frac{\partial^2 x G_A( y,\xi)}{\partial y \partial \xi}\,\,=\,\,
\frac{2\,R^2_A \,\,Q^2}{ \pi^2}\,\,\{\,\,C\,\,+\,\,\ln \kappa\,\,+
\,\,E_1 ( \kappa )\,\,\}\,\,.
\eeq
Rewriting \eq{102} in terms of $\kappa$  given by
\beq \label{KAPPA}
\kappa \,\,=\,\,\frac{N_c \as \pi }{2 Q^2 R^2_A}\,x G_A(x,Q^2)
\eeq
we obtain:
\beq \label{103}
\frac{\partial^2 \kappa( y,\xi)}{\partial y \partial \xi}\,\,+\,\,
\frac{\partial \kappa(y, \xi)}{\partial y}\,\,=\,\,
\frac{ N_c\, \as\,}{\pi}\,\,\{\,\,C\,\,+\,\,\ln \kappa(y, \xi)\,\,+
\,\,E_1 ( \kappa(y, \xi) )\,\,\}\,\,\equiv\,\,F(\kappa)\,\,.
 \eeq
Now, let us consider the expression of \eq{103} as the equation for  $\kappa$
This equation has the following nice properties:

1.  It  sums all contributions of the order $ (\,\as\,y\,\xi\,)^n$ 
absorbing them in $x G_A (y,\xi)$, as well as all contributions of the order
of $\kappa^n$.
Therefore, this equation solves the old problem, formulated in Ref.\cite{GLR}
and 
 for $N_c\,\rightarrow \,\infty $ \eq{103} gives the complete
solution to our problem, summing all SC;

2 .The solution of this equation matches with the solution of the DGLAP
 evolution equation in the DLA of perturbative QCD at $\kappa\,\rightarrow \,0$;

3.  At small values of $\kappa$ ( $\kappa\,<\,1$ ) 
 \eq{103} gives the GLR equation. Indeed, 
for small $\kappa$ we can expand the r.h.s of \eq{103} keeping only the
 second term. Rewriting the equation through the gluon structure function
 we have
\beq \label{GLR}
\frac{\partial^2 x G_A( y,\xi)}{\partial y \partial \xi}\,\,=\,\,
\frac{\as N_c}{\pi}\,x G(x,Q^2)\,\,-\,\,\frac{\as A^2}{2 R^2_A}\,( x G(x,Q^2))^2
\,\,,
\eeq
which is the GLR equation \cite{GLR}
 with the coefficient in front of the second term
 calculated by Mueller and Qiu \cite{MUQI}.

4. For $\as y \xi \,\approx\,1$ this equation gives the Glauber ( Mueller ) 
formula, that we have discussed in details.

5. This equation almost coincide with the equation that L.Mclerran
with collaborators \cite{MCLER} derived from quite different approach
and with different technique. We are sure that {\it almost} will
disappear when they will do more careful averaging over transverse distances.

Therefore, the great advantage of this equation in comparison
 with the GLR one is the fact that it describes the region of large $\kappa$
and provides the correct matching both with the GLR equation and with the
Glauber ( Mueller ) formula.

Eq. (\ref{103}) is the second order differential equation in partial derivatives
and we need two initial ( boundary ) conditions to specify the solution.
The first one is obvious, namely, at fixed  $y$ and $Q^2 \,\rightarrow \,\infty$
$$
\kappa\,\,\rightarrow\,\,\frac{N_c \,\as\,\pi \,A}{ 2\,Q^2\,R^2_A}
\,x G^{GLAP}_N (x,Q^2)\,\,.
$$
The second one we can fix in the following way: at $ x = x_0 \,\,(y = y_0)$
which is small, namely, in the kinematic region where  $\as y \xi\, \leq\, 1$
\beq \label{104}
\kappa\,\,\rightarrow\,\,\kappa_{in}\,\,=\,\,
\frac{N_c \,\as\,\pi }{ 2\,Q^2\,R^2_A}
\,x G_A (x,Q^2)\,\,,
\eeq
where $x G_A$ is given by the Mueller formula ( see \eq{MF}).
Practically, we can take $x_0 \,=\,10^{-2}$, because  corrections to the MF
are  small at this value of $x=x_0$.

{\em 5.3 The asymptotic solution.}

First observation is the fact that \eq{103} has a solution which depends only
 on $y$. Indeed, one can check that $\kappa \,=\,\kappa_{asymp}(y)$ is
the solution of the following equation:
\beq \label{105}
\frac{d \kappa_{asymp}}{d y}\,\,=\,\,F(\,\kappa_{asymp}\,)\,\,.
\eeq
The solution to the above equation is: 
\beq \label{106}
\int^{\kappa_{asymp}(y)}_{\kappa_{asymp}(y=y_0)}\,\,
\frac{ d \kappa'}{F(\kappa')}
\,\,=\,\,y - y_0\,\,.
\eeq

It is easy to find the behavior of the solution to \eq{106} at large value
of $ y$ since $F(\kappa)\,\,\rightarrow\,\,\bar \as \ln \kappa $ at
large $\kappa$ ( $\bar \as = \frac{N_c}{\pi}\,\as$ ). It gives
\beq \label{107}
\kappa_{asymp}\,\,\rightarrow \,\,\bar \as y \,\ln(\bar  \as y)\,\,\,\,\,\,
at\,\,\,\,\,\,\,\bar \as y \,\,\gg\,\,1\,\,.
\eeq
At small value of $y$, $F(\kappa)\,\,\rightarrow\,\,\bar \as \kappa$ and
we have:
\beq \label{108}
\kappa_{asymp}\,\,\rightarrow\,\,\kappa_{asymp} ( y = y_0 )
\,\,e^{\bar \as ( y - y_0)} \,\,.
\eeq
The solution is given in Fig.\ref{asy} for $\bar \as = 1/4$  in the whole region
 of $y$ for different nuclei in comparison with our calculations based
on the MF. We chose the value of $\kappa_{asymp} (y =y_0)$ from \eq{104}.
We claim this  solution is the asymptotic solution to \eq{103} and  will 
argue on 
this point a bit later.

 For nuclei the SC incorporated in the
asymptotic solution turn out to be much stronger than 
the SC in the Glauber approach for any $Q^2 \,>\,1\,GeV^2$ at $x\,>\,10^{-2}$.
In this kinematic region  the solution of \eq{103} is drastically
different from the Glauber one.

A general conclusion for Fig.\ref{asy} is very simple:
the amount of  shadowing which was taken into account in the MF
 is not enough , at least for the gluon structure function in nuclei at
$x\,<\,10^{-2}$ and we have to solve \eq{103}
to obtain the correct behavior of the gluon structure function for nuclei.

\begin{figure}[hptb]
\begin{tabular} {c c}
 \epsfig{file=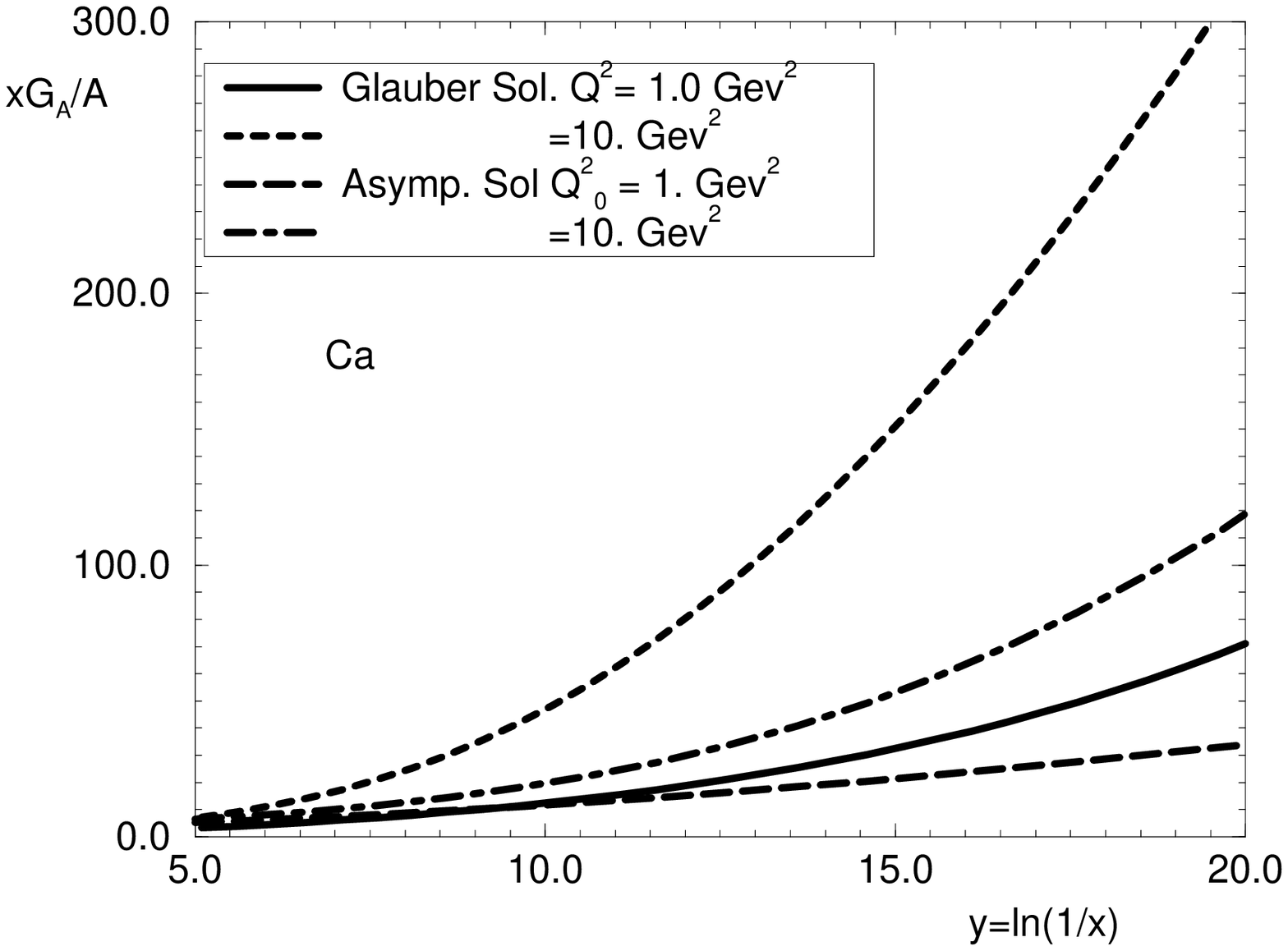,width=70mm} &
\epsfig{file=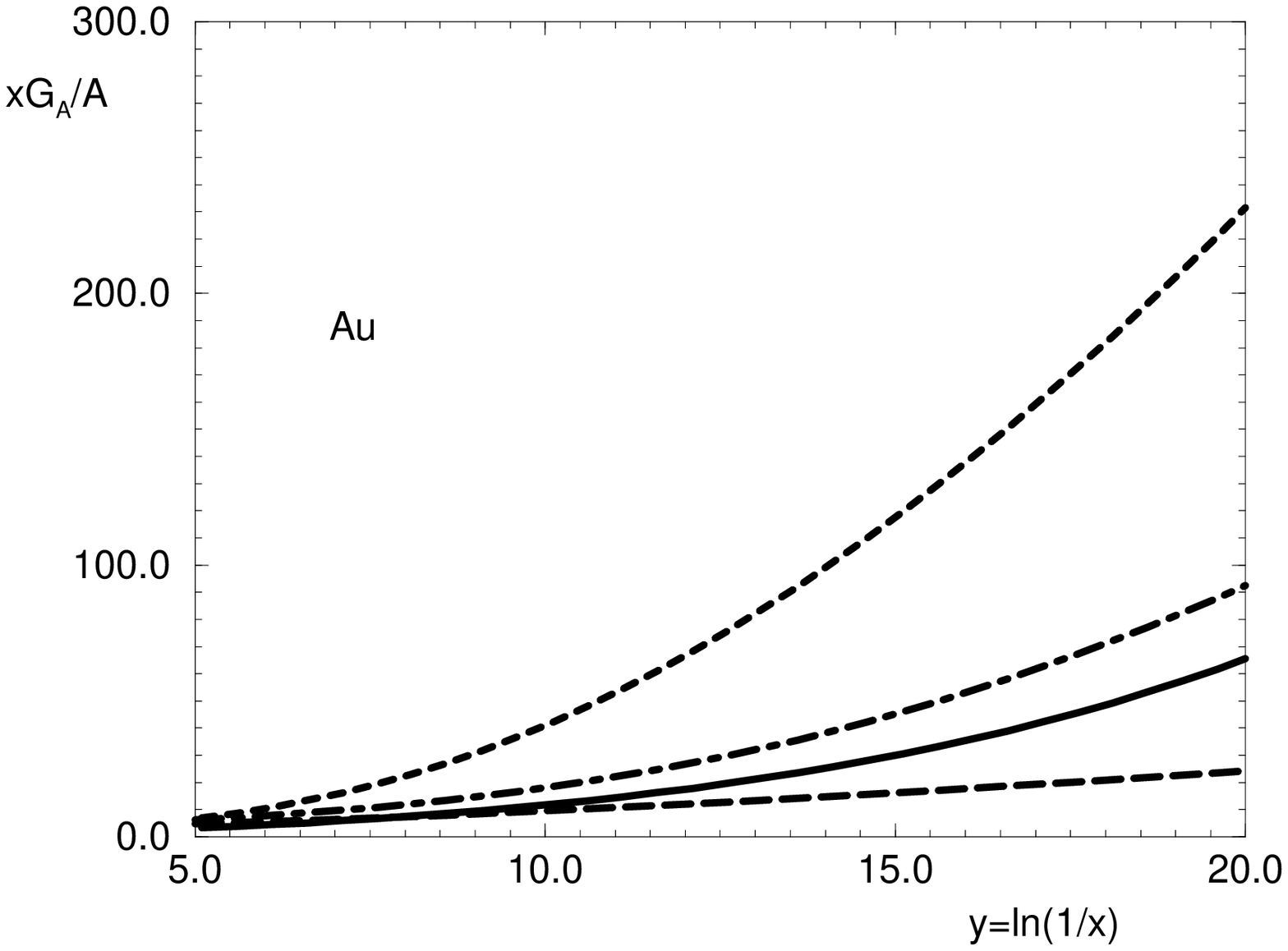,width=70mm}  \\
\end{tabular}
\caption{ \em The Glauber approach  and asymptotic solution for
different nuclei.}
\label{asy}
\end{figure}

Now, we would like to show that solution \eq{103} is the asymptotic solution
 of the new evolution equation.
 In order 
to check this we need to prove that this solution is stable.
It means that if  add a small function $\Delta \kappa$ and searching for
 the solution to the equation in the form
$\kappa \,=\,\kappa_{asymp} \,+\,\Delta \kappa$, we have to prove that
$\Delta \kappa $ turns out to be small.namely, $\Delta \kappa\,\ll\,\kappa$.
The following linear equation can be written for $\Delta \kappa$:
\beq \label{109}
\frac{\partial^2 \Delta \kappa( y, \xi  )}{\partial y\,\partial \xi}\,\,+\,\,
\frac{\partial \Delta \kappa( y, \xi}{\partial y}\,\,=\,\,\frac{d F(\kappa)}{
d \kappa}\,|_{\kappa = \kappa_{asymp}(y)}\,\,\Delta \kappa (y, \xi)\,\,.
\eeq
In Ref.\cite{AGL} was proven, that the solution of \eq{109} is much smaller
 than $\kappa$.

Therefore the asymptotic solution has a chance to be the solution of our
 equation in the region of very small $x$. To prove that the asymptotic
 solution is the solution to the equation we need to solve our equation 
in the wide kinematic region starting with our initial condition.
 We managed to do this only in semiclassical approach.

{ \em 5.4 Semiclassical Approach.}

The semiclassical approach has been adjusted
to the solution of the nonlinear equation of eq.(\ref{103})-type in Refs.
\cite{GLR,Collins90,Bartels91} ( for simplicity, we assume that $\as$ is 
fixed ).

In the semiclassical approach we are looking for the solution of eq.
(\ref{103}) in the form
\bea
\kappa = e^S
\eea
where $S$ is a function with partial derivatives: $\frac{\partial S}{
\partial y} = \o $ and $\frac{\partial S}{\partial \xi} = \gamma $ 
which are smooth function of $y$ and $\xi$.
It means that
\bea
\frac{\partial^2 S}{\partial \xi \partial y} \ll \frac{\partial S}{
\partial y} \cdot \frac{\partial S}{\partial \xi} = \o \gamma
\label{118}
\eea

Using eq.(\ref{118}), one can easily rewrite eq.(\ref{103}) in the form
\bea
\frac{\partial S}{\partial y} \frac{\partial S}{\partial \xi} +
\frac{\partial S}{
\partial y} = e^{-S} F(e^{S}) \equiv \Phi(S)
\label{119}
\eea
or
\bea
\o (\gamma + 1) = \Phi (S)
\label{120}
\eea

We are going to use the method of characteristics( see, for example,
 ref.\cite{Sneddon}).
 For equation in the form
\bea
F(\xi, y, S, \gamma , \o ) = 0
\label{121}
\eea
we can introduce the set of characteristic lines $ (\xi(y), S(y), \o (y),
\gamma (y) ) $, which satisfy a set of well defined equations (see, for example, Refs. \cite{Collins90} \cite{Bartels91} for the method and Ref  \cite{AGL} for detailed calculation).
Using eq.(\ref{119}) and  eq.(\ref{120}),
 we obtain the following set of equations for the characteristics:
\bea
\frac{d \xi}{d y} = \frac{\Phi(S)}{(\gamma +1)^2}  \, ; \,\,\,\, 
\, \frac{d S}{d y} =   \frac{2 \gamma + 1}{(\gamma +1)^2}\Phi (S) \, ; \,\,\,\, 
\frac{d \gamma}{d y} = \Phi'_{S} \frac{\gamma}{\gamma +1} \, \,,
\label{125}
\eea 
where $\Phi'_S\,=\, \frac{\partial \Phi}{\partial S}$.
The initial condition for this set of equations we derive from eq.(\ref{104}),
namely
\bea
S_0 = ln \kappa_{in} (y_0, \xi_0) \nonumber \\
\gamma_0  = \left. \frac{\partial ln \kappa_{in} (y_0 , \xi )}
{\partial \xi} \right|_{\xi = \xi_0}
\label{129}
\eea

The main properties of these equations have been considered in Ref.\cite{AGL}
analytically, however, here, we restrict ourselves mostly the numeric solution
of these equations.

We set the  initial condition  $y = y_0 = 4.6$ ($ x_B = 10^{-2}$), 
where the shadowing correction is not big and the evolution starts from
$\gamma < 0$. In this case  $d \gamma / d y > 0$ and the value of 
$\gamma$ increases . At the same time $d S / d y < 0$ and $S$ decreases if
 $\gamma_0 < -1/2$. With the decrease of $S$, the value of $\Phi'_{S}$
becomes smaller and after short  evolution the trajectories of the nonlinear 
equation start to approach the trajectories of the DGLAP equations. We face
this situation for any trajectory with $\gamma_0 $ close to -1. If the 
value of $\gamma_0$ is smaller than $- \,\frac{1}{2}$ but the value of
$S_0$ is sufficiently big, the decrease of $S$ due to evolution cannot
provide a small value for $\Phi'(S)$ and $\gamma$ increases until its value
becomes bigger than $ - \frac{1}{2}$ at some value of $y= y_c$. In this case
 for $y > y_c$   the trajectories behave as in the case with $\gamma_0 \,>
 \,-\,\frac{1}{2}$.
 For
$\gamma_0 > -1/2$, the picture changes crucially. In this case, $d S/ d y
> 0 $ , $d \gamma / d y > 0$  and both increase. Such trajectories go apart 
from the trajectories of the DGLAP equation and nonlinear effects play more
and more important role with increasing $y$. These trajectories approach 
the asymptotic solution very quickly.

For the numerical solution we use the 4th order Runge - Kutta method
to solve our set of equations with the initial distributions of  \eq{129}.
 The result of the solution is given in
Figs.\ref{scn} and \ref{sca}. 
In these figures we plot the bunch of the trajectories
with different initial conditions. For the nucleon ( Fig.\ref{scn} )
 we show also
the dependence of $ \gamma$ along these trajectories. One can notice 
that the trajectories behave in the way which we have discussed in our
qualitative analysis. It is interesting to notice that the trajectories,
which are different from the trajectories of the GLAP evolution equations,
 start  at $y = y_0 = 4.6$ with the values of $Q^2$ between $0.5 GeV^2$ and 
$2.5 GeV^2$ for a nucleon. It means that, guessing  which is the boundary
condition   at $Q^2 = Q^2_0 = 2.5 GeV^2$,  we can hope that the 
linear evolution equations ( the DGLAP equations) will describe the evolution
of the deep inelastic structure function in the limited but sufficiently
wide range of $Q^2$.

In Figs. \ref{scn} and \ref{sca} we plot 
also  the lines with definite value of the ratio
$R\, = \,\frac{xG(x,Q^2)( generalized\,\,\, equation)}{x G(x,Q^2) (GLAP)}$
 (horizontal lines). These lines give the way to estimate how big are the SC.
One can see that they are rather big.
\begin{figure}[hptb]
\begin{center}
\begin{tabular}{ c c}
\epsfig{file=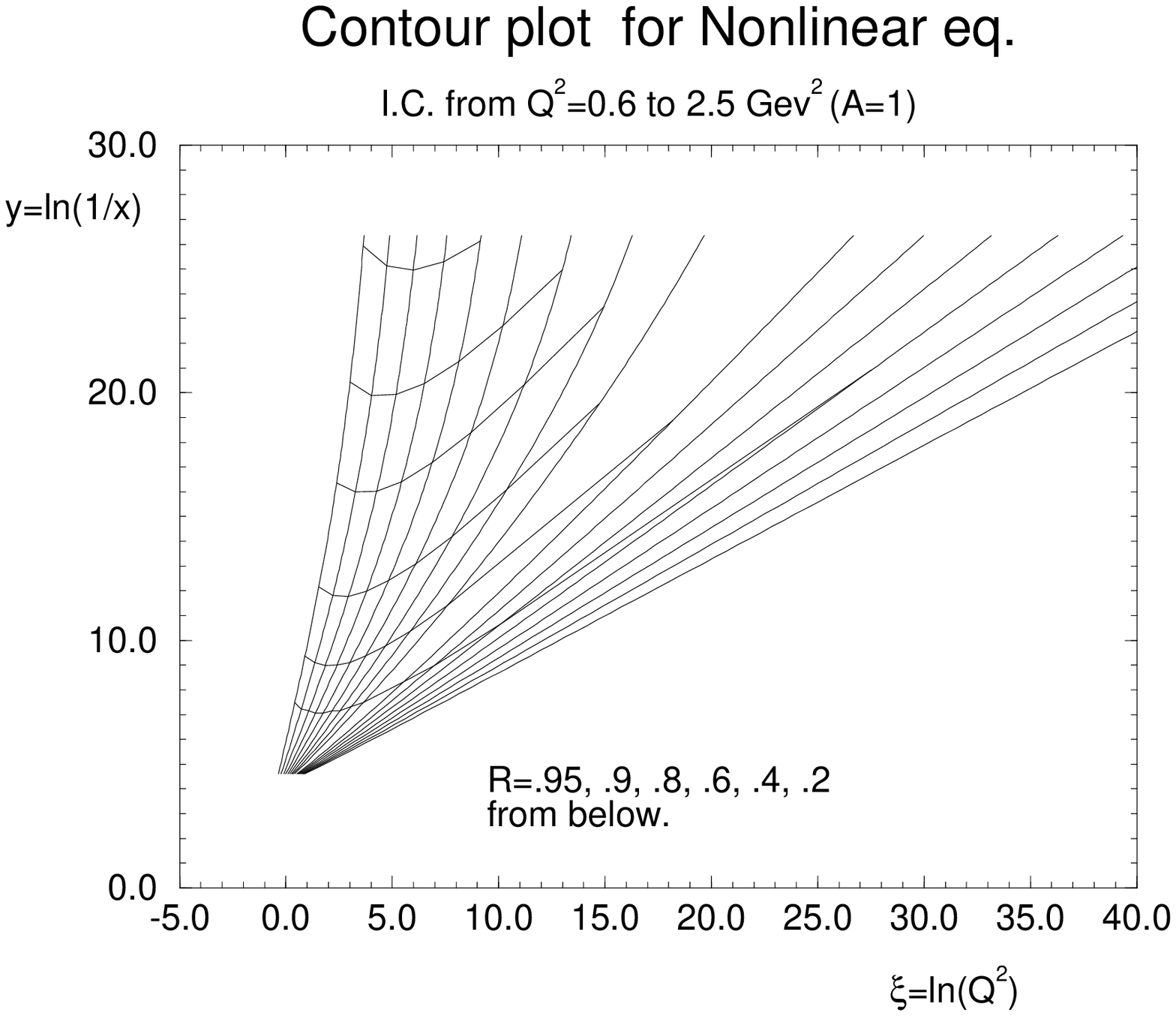,width=70mm} & \epsfig{file=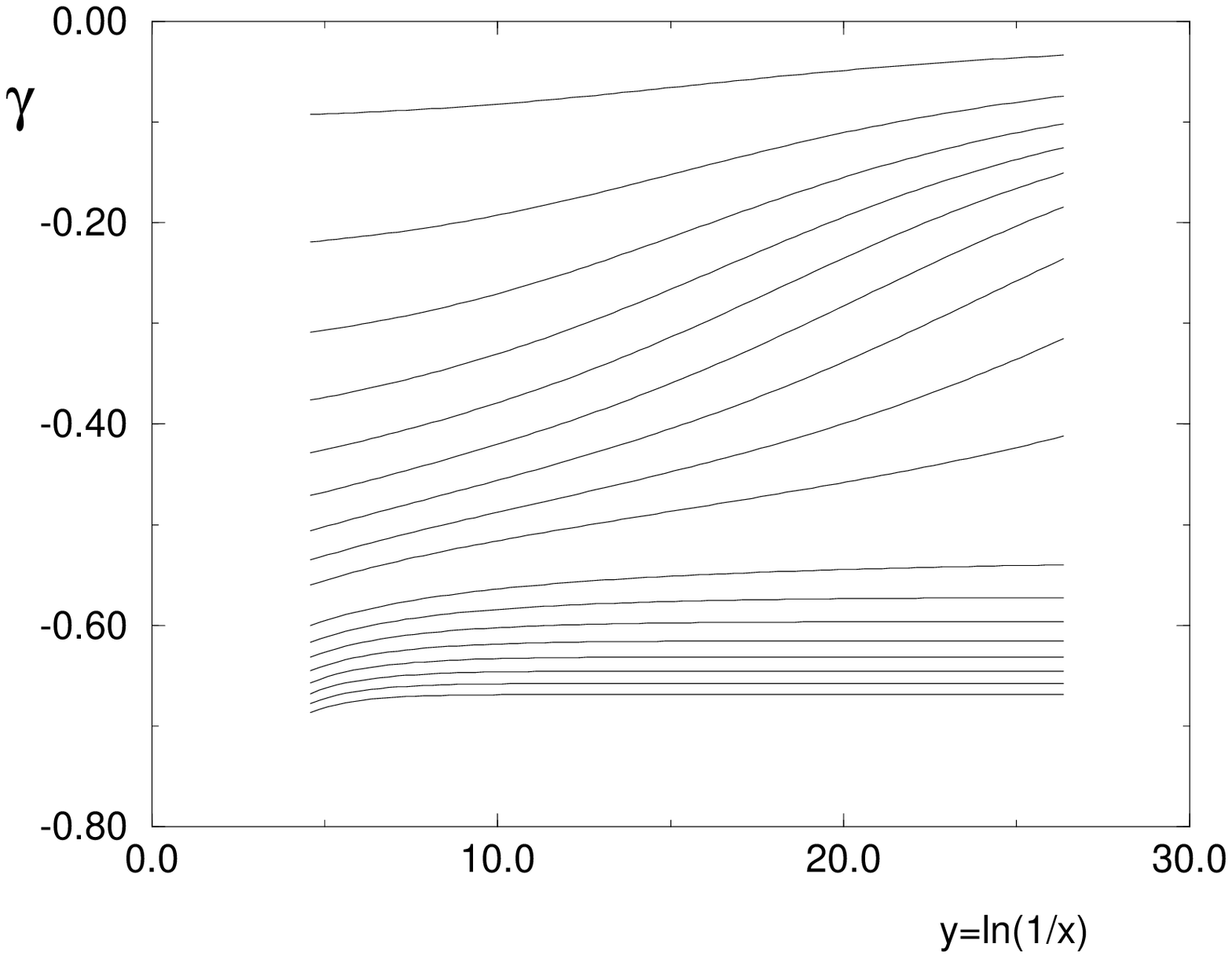,width=70mm}\\
\end{tabular}
\end{center}
\caption{ \em The trajectories and contour plot for the solution of 
the generalized
 evolution equation for N.\,\,\,$R\, = 
\,\frac{xG(x,Q^2)( generalized\,\,\, equation)}{x G(x,Q^2) (GLAP)}$.}
\label{scn}
\end{figure}
\begin{figure}[hptb]
\begin{center}
\begin{tabular}{ c c}
\epsfig{file=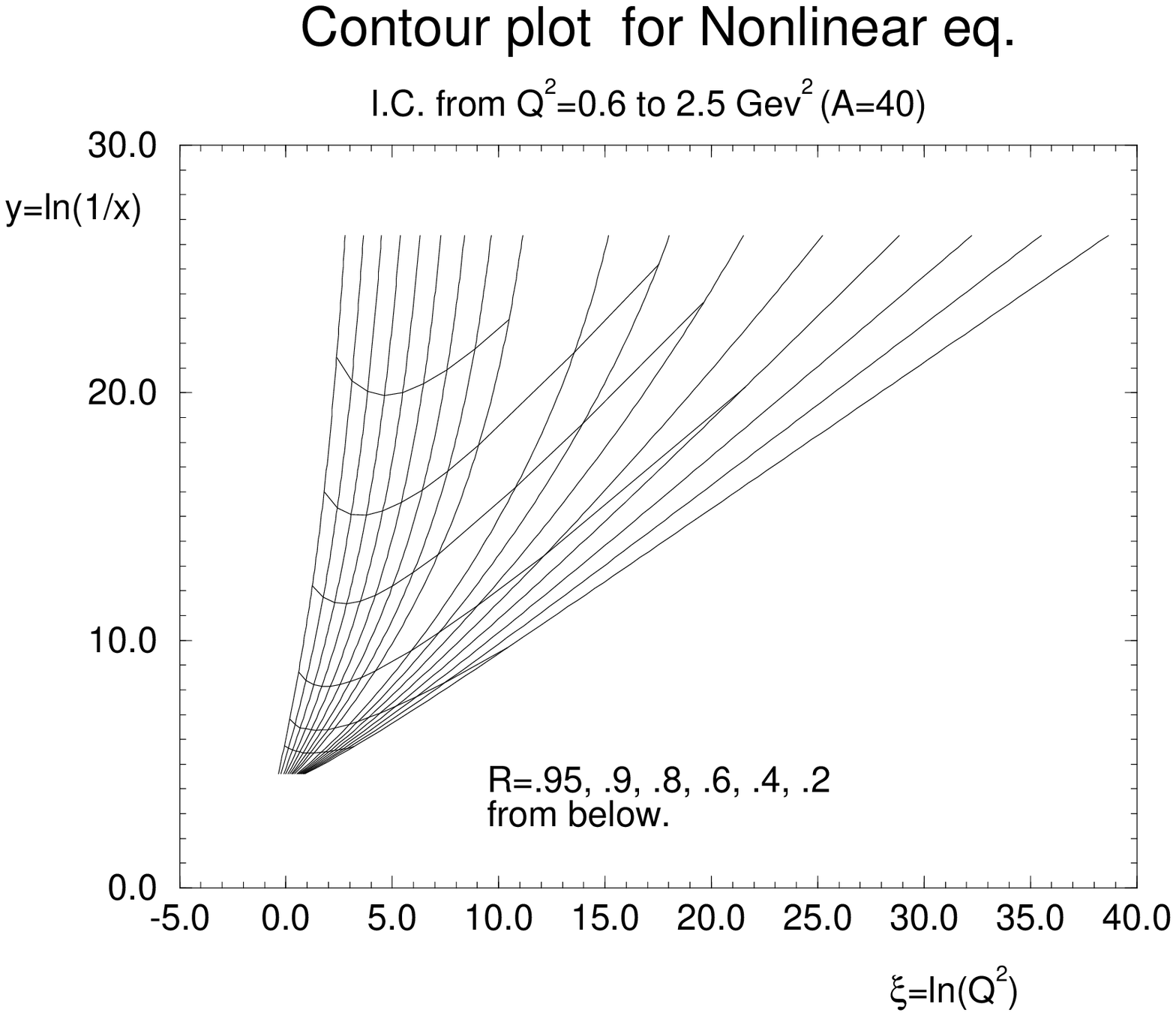,width=70mm} & \epsfig{file=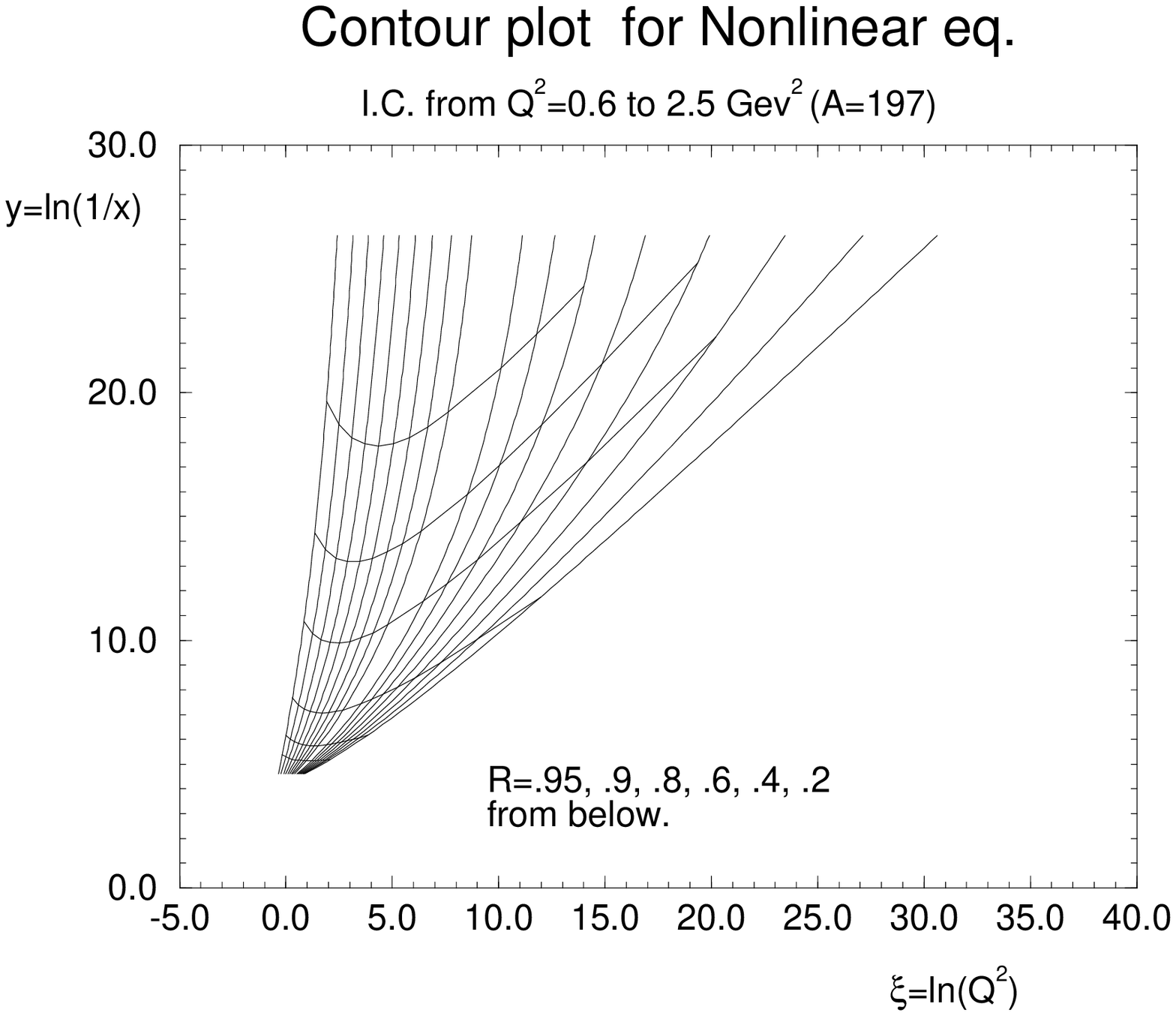,width=70mm}\\
\end{tabular}
\end{center}
\caption{\em The trajectories and  contour plot for the solution of the generalized
 evolution equation for Ca and Au.\,\,\,$R\, = 
\,\frac{xG(x,Q^2)( generalized \,\,\,equation)}{x G(x,Q^2) (GLAP)}$.}
\label{sca}
\end{figure}

We have discussed only the solution with fixed coupling constant which we
put equal to $\as = 0.25$ in the numerical calculation. The problem how to
solve the equation with running coupling constant is still open.

{\em 5.5  The generalized evolution equation versus the GLR equation.}

In Ref.\cite{AGL} we studied in detail the solution to the GLR equation in
 the same semiclassical approximation. Our conclusion is that the GLR
 equation gives much stronger SC than the generalized evolution equation.
 This difference we can see comparing  the solution to the both equation in the
region ultra small $x$.

Indeed, our asymptotic
solution turns out to be quite different from the GLR one. The GLR solution
in the region of very small $x$ leads to saturation of the gluon density
\cite{Collins90,Bartels91,BALE}. Saturation means that $\kappa$ tends
to a constant in the region of small $x$. The solutions of
\eq{103} approach the asymptotic solution  at $ x\,\rightarrow\,0$, which
 does not depend on $Q^2$, but exhibits sufficiently strong dependence
 of $\kappa$  on
$x$ ( see Fig.\ref{asy} ), namely $\kappa \,\propto\,\as \ln (1/x)\ln\ln(1/x)$.
The absence of saturation does not contradict any physics since gluons
are bosons and it is possible to have a lot of bosons in the same cell of
 the phase space. We should admit that A. Mueller first came to the same
 conclusion using his formula in Ref.\cite{MU90}.

\section{Next  steps.}

Here, we list our problems that have to be solved to complete our study of
 the SC :

1. Calculation of $F^A_2(x,Q^2)$ to compare our calculation of the SC with 
the available experimental data.

2. Recalculation of the SC using more reliable Wood-Saxon parameterization
 for profile function $S(b_t)$ instead of the Gaussian one. The form of
 the profile function especially essential to obtain a reliable estimates for
the SC in the region of the moderate $x\,\leq \, \frac{1}{2 m R_A}$.

3. Solution of the generalized evolution equation for running $\as$.
The experience of solving the GLR equation tells us that there is a principal
difference in the solutions for fixed and running $\as$, namely, the critical
 line of the GRL equation appears only for running $\as$ \cite{GLR}.
We think that it is very important to study  the generalized equation with
 running $\as$ and to compare this solution with the solution of the 
GRL equation.
  
4. We have discussed that for RHIC energies it is very important to study
in more details the effect of the final life-time of th gluon in a nucleus.
We plan to recalculate the SC replacing $xG(x,Q^2)$ in our formulae by 
$xG(x,Q^2,q_z)$ for which  the kernels of the evolution equations have
 been calculated in Ref.\cite{GLR}.

5. In all our calculations we neglected the parton interaction inside
$ GG + N$ scattering. Our estimates, which have been presented in section 2,
shows that this interaction should be very important. Indeed, for example, 
in the Mueller formula we have to change the parameter $\kappa_G$ due to
the parton interaction inside the nucleon. This change is simple, the only that
we need to do is to replace the number of collisions $A/\pi R^2_A$ by
$$
\frac{A}{\pi R^2_A}\,\,\rightarrow\,\,\frac{A}{\pi R^2_A}\,\,+
\,\,\frac{1}{\pi R^2_N}$$
in the definition of $\kappa_G$ in \eq{MF}. It means that all results will
 be the same but nucleus with the new effective number of nucleons:
$$
A^{\frac{1}{3}}_{eff}\,\,=\,\,A^{\frac{1}{3}}\,\,+\,\,\frac{R^2_0}{R^2_N}\,\,,
$$
where $R_A = R_0 \,A^{\frac{1}{3}}$. Using our estimates for
 $R^2_N = 5 GeV^{-2}$ we can see that effective A for the gold is 
$
A^{\frac{1}{3}}_{eff} \,\,=9.6 $ instead $A^{\frac{1}{3}} \,=\,6 $.
For light nuclei the change is even more essential.
Therefore, we are planning to take into account  the parton interaction
 inside a nucleon as soon as possible.

6. We have neglected all correlations between partons of the order
 $\frac{1}{N^2_c}$ which could be sizable in the case of the nucleus DIS.
We suppose to study this problem using the technique that has been developed
 in Ref.\cite{LALE95}.

7. Everywhere through the paper we used the DLA of perturbative QCD.
However, the key assumption that simplify our theoretical approach was
the $\as \ln(1/x) \,\approx\,1$ approximation. We plan to develop our
approach in the case of the BFKL dynamic and, therefore, to get rid of 
our assumption that $\as \ln (Q^2/Q^2_0) \,\,\approx\,\,1$. We consider this 
generalization as an important  step, since our result that we have no
 saturation of the gluon density in nuclei even at ultra small $x$ could be
an artifact our double log approximation of perturbative QCD.

\section{And what?}

We presented here our approach to the SC and a natural question arises:and what?
What and how we can do for the RHIC physics. How our approach can help in
creating of the reliable Monte Carlo code for nucleus - nucleus interaction
 at high energies.We are going to answer these hot question in this section.

Let us consider first the space time structure of the nucleus - nucleus
 interaction ( see Fig.\ref{fig.7.1}).
\begin{figure}
\vspace{4.5cm}
\includegraphics{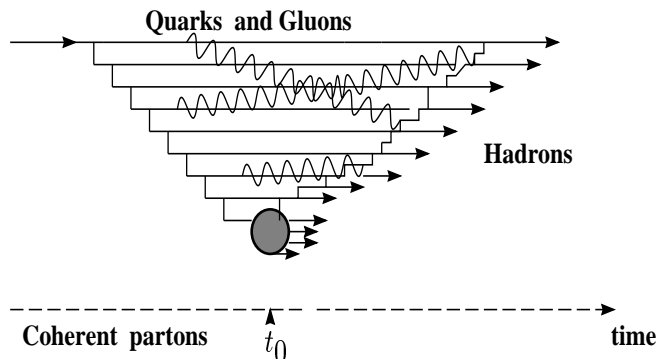}
\vspace{-1cm}
\caption{{\em Four stages of nucleus - nucleus collision.}}
\label{fig.7.1}
\end{figure}
One can see  four stages of this process:

1. For time smaller that $t_0$, where $t_0$ is the time of the first
 parton - parton interaction, we have a very coherent system of parton,
confined in our both nuclei. We know almost nothing about this system.

2. At time $t_0$ the first parton - parton interaction occurs and we believe
that this interaction destroys the coherence of our parton system at 
the very instant.

3. During time from $t_0$ till $t_h$, where $t_h$ is the hadronization time, 
we have a quark - gluon stage of the process. We believe that we can reach
a simple and economic understanding this stage in framework of QCD. 
We also believe that new collective phenomena could be created in
 the nucleus - nucleus interaction during this stage of the process such as
 the Quark - Gluon Plasma mostly because of the high density of the
 produced gluons.
For 
this stage we have the Monte Carlo codes based on QCD, the lattice calculation
and a lot of beautiful ideas that has been discuss at this conference.

4. The last stage - hadronization is a black box. Nothing is known, but the
success of the Local- Hadron-Parton Duality in the description of the LEP data
allows us to hope that this stage could be not very important for our
 understanding of the nucleus - nucleus collisions.

Our approach can define the initial condition at $t=t_0$ for the third stage.
What can we do?

1. We are able to calculate the inclusive cross section for gluons at $t=t_0$
 or, in other words, define the gluon distribution at $t=t_0$. Actually,
it has been done by C.Escola \cite{ESCOLA}
and his collaborator  and has been presented at
 this conference. We can only improve his treatment of the SC which was based
 on the GLR equation. However, let us discuss briefly the formula for
the inclusive gluon cross section. It can be written using the factorization
 theorem \cite{FACTOR} in the form:
$$
\frac{d \s}{d y \,d p^2_t}\,\,=\,\,\propto\,\,
\int \, d x_1 d x_2\,x_1 G_{A_1}( x_1,p^2_t)\,x_2G_{A_2}(x_2,p^2_t)\,\,
\frac{\as}{p^4_t}\,\,
$$
where the last factor is the hard gluon - gluon cross section and $y$ and
 $p_t$ are rapidity and transverse momentum of produced gluon, respectively.
One can see that this cross section is infrared unstable and diverges at small 
values of $p_t$. The SC provides a natural scale that cut off this divergence.
A rough estimate for this new scale can be done from equation
$$
\kappa_G (x,r^2_{\perp} = \frac{1}{Q^2_0(x)})\,\,=\,\,1\,\,
$$
(see Fig.4 ).
For $p_t\,<\,Q_)(x)$ the gluon structure function $xG(x,p^2_t) \,\,\propto\,\,
p^2_t$ and one can see that the number of gluon with transverse momenta smaller
than $p_t = Q_0(x)$ turns out to be very small.

2. We can calculate also the double inclusive cross section which gives
the two gluon correlation function at $t=t_0$. We would like to stress that for
nucleus - nucleus collision this correlation function is big and have to be
taken into account. Indeed, we have two different
contribution to the double inclusive process, pictured in Fig.\ref{fig.7.2}:
the production of two gluons from one parton cascade (see Fig.\ref{fig.7.2}a)
and from two parton cascades ( see Fig.\ref{fig.7.2}b ).
However, for nucleus - nucleus collisions the first contribution is proportional
to $A_1 A_2$ ( without the SC) while the second is much bigger and it is of
 the order of 
$ A^2_1 A^2_2 \,\frac{R^2_{A_1} \,+\,R^2_{A_2}}{\pi R^2_{A_1} R^2_{A_2}}$
( without the SC and for the Gaussian profile function). Using our approach we
can calculate the two gluon correlation function within better acuraccy than
the above simple estimates. We hope, that these two observables: gluon 
distribution and two gluon correlation function will be enough for reliable
description of the initial condition for the QCD motivated cascade  during
the third stage of our process.

\begin{figure}
\vspace{6cm}
\includegraphics{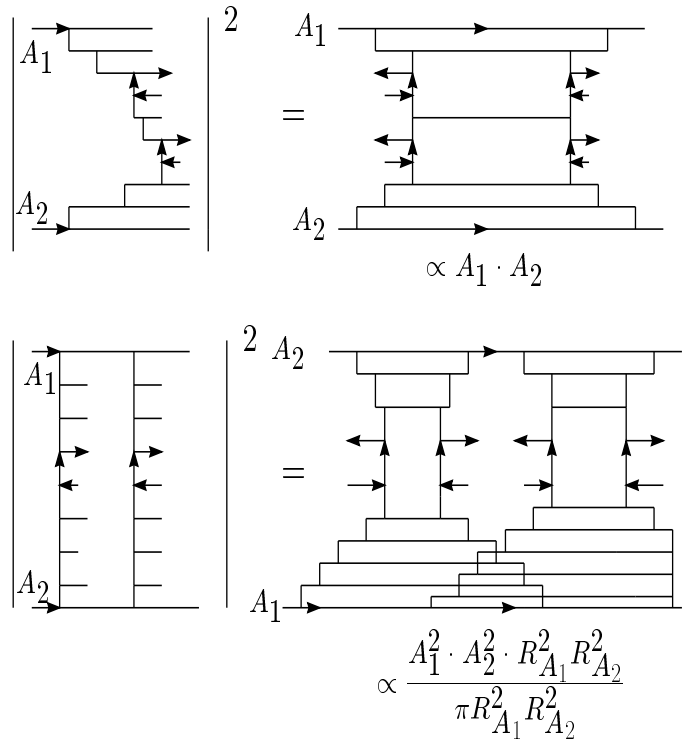}
\vspace{4cm}
\caption{{\em Double inclusive cross section in QCD.}}
\label{fig.7.2}
\end{figure}

3. We think that these two observables: multiplicity of gluons and two
 gluon correlation will be enough to define the initial condition for
current Monte Carlo codes. However, we think that these codes are doing
something wrong. Indeed, we learned from A. Mueller \cite{MU94} that correct 
degrees of freedom  for  parton cascading looks in the simplest way
and which could be used  for a probabilistic interpretation and therefore,
they are natural degrees of freedom for Monte Carlo simulations are not quark
 and gluons but colourless quark - antiquark dipoles. The gluon structure
function is the probability to find a colourless dipole with the size $r_{\perp}
\,\geq\,\frac{1}{Q}$. Therefore, we think that the code should be written 
for such dipoles and their interaction. We shall answer the questions:(i)
how to calculated the average multiplicity of dipoles with the size $r_{\perp}$
and (ii) how to calculate the correlations between such dipoles.
We are going to do this in the nearest future.

4. Now we want to discuss a hot question how to mix the "soft" and "hard"
Pomerons. The common way of doing such a mixture is to use the Glauber formula
and replace in this formula $\s(r^2_{\perp}\,\rightarrow
\s_{soft} \,+\,\s_{hard}$. We think this is a correct procedure 
to obtain an estimate how important soft or/and hard processes. In section
 3 we argued that this is the  most economic way of doing which satisfies
the $s$-channel unitarity. However, all Monte Carlo programs that we know
use for the calculation of $\s_{hard}$ the factorization formula, namely
$$
\s_{hard}\,\,=\,\,\frac{!}{2}\,\int_{p^2_0}\,d p^2_t \,x_1G(x_2,p^2_t)
x_2G(x_2,p^2_t) \,\frac{\as^2}{p^4_t} \,\,,$$
which describes really the inclusive production of gluons. The factor 1/2
in front does not help because to find $\s_{hard}$ we need to calculate the 
real multiplicity but not the number of gluon line in the Feynman diagram.
In Fig.\ref{fig.7.3} we picture the $\as^3$ corrections to the hard cross
 section considering the scattering of two mesons made from heavy quarks.
Perturbative QCD is certainly a good tool to study such processes.
From this picture one sees that including the inclusive cross section in the
place of the total we missed the radiative correction to the the partial cross
section with four quarks i the final state.

\begin{figure}
\vspace{4.5 cm}
\includegraphics{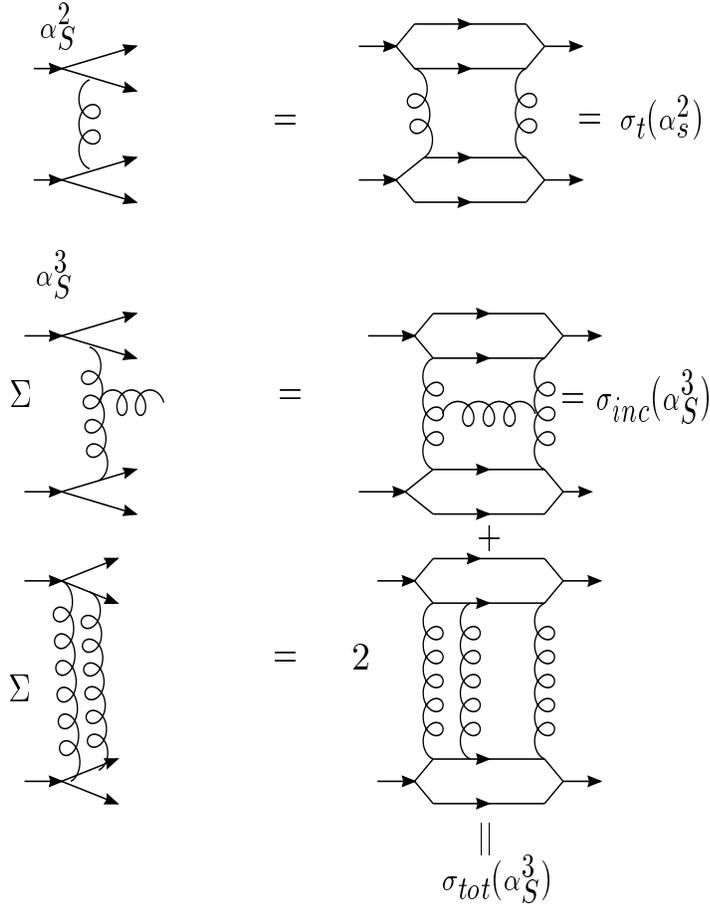}
\vskip 200 pt
\caption{{\em Total cross section in $\as^2$ and $\as^3$ orders of perturbative QCD.}}
\label{fig.7.3}
\end{figure}

Our way of doing is the following. We will write the Mueller formula or
our more sofisticated approach for dipole (with size $r_{\perp}$
 scattering with a nucleus. To find the proton - nucleus cross section
we need to calculate the integral:
$$
\s(p A)\,\,=\,\,\int^1_0 \prod^3_1  d z_i \int  \prod \frac{d^2 r^{ik}_{\perp}}
{2 \pi} \Psi_p ( z_i,r^{ik}_{\perp}\,\,
 2 \,\{\,\,1\,\,-\,\,e^{ -\,\frac{1}{2} \s_n(r^{ik}_{\perp } S(b_t)}\,\,\}
\,\,\Psi^*(r^{ik}_{\perp},z_i)\,\,.
$$
To find the wave function of the nucleon we have to use a model,for
example the constituent quark model or instanton liquid model. The nice
feature of this formula that the typical $r^{ik}_{\perp}$ will be of the
order 1 $GeV^{-1}$ due to the SC. It means that we need to know the wave 
function at sufficiently small distances where we have some control from 
lattice calculations and QCD sum rules.
 This formula takes into account correctly hard process and give the 
factorization formula for the inclusive production. We suppose to do
an estimates using the model for the nucleon wave function. If they will show
that we need some admixture of the soft processes we will add to $\s_N$ in
the above formula in an usual phenomenologic way, using the model of, so called,
soft Pomeron.

\section{Conclusions.}
We have two conclusions:

1. We hope that we convinced you that we are on the way from our Really
Highly Inefficient Calculation to your RHIC. Much work is need to clarify
the initial condition for the QCD phase of nucleus - nucleus interaction
and this is the first and the most important task which we need to attack,
since it will determine the correct degrees of freedom for further
evolution of QCD cascades.

2. Everything that we have talked about satisfies the third law of
 theoretical physics:{\it Any model is a theory which we apply to a kinematic
region, where we cannot prove that this theory is wrong}.
We firmly believe that correct SC will provide the picture of
 the nucleus - nucleus interaction in which hard and semihard processes will
 play a crucial role with only small if any contamination
 of the soft contribution.

{\bf Acknowledgements:} 

One of us (E.M.L.) is very grateful to Sid Kahana
 for creation a stimulating atmosphere of discussion at
RHIC'96 Workshop and for enlighting discussions of the
 difficult theoretical problems in the  modeling heavy ion collisions.
E.M.L. thanks all participant of the hard working group and especially 
 Yu. Dokshitzer,C.Eskola, A. Mueller and M.Strikman for fruitful
 discussions on the subject of the talk and related topics.
 MBGD thanks A. Capella and D.Schiff for enlightening discussions.
 Work partially
 financed by  CNPq, CAPES and FINEP, Brazil.

\end{document}